\documentclass[aps,prl,twocolumn,groupedaddress,preprintnumbers]{revtex4-1}

\usepackage{lipsum}
\usepackage{amsmath}	
\usepackage{bm}		
\usepackage{amssymb}	
\usepackage{color} 		
\usepackage{graphicx} 	
\usepackage[T1]{fontenc}
\usepackage{times} 		
\usepackage{mathrsfs} 
\usepackage{ulem}
\usepackage{hyperref} 
\hypersetup{
    colorlinks=true,       
    linkcolor=blue,        
    citecolor=blue,        
    filecolor=magenta,     
    urlcolor=blue          
}
\usepackage[all]{hypcap} 

\newcommand{\WD}{{\text{WD}}}
\newcommand{\EM}{\mathrm{em}}
\newcommand{\dvec}{{\bm \nabla}}
\newcommand{\Avec}{{\bm A}}
\newcommand{\dAvec}{\delta \hspace{-0.04cm} \Avec}

\newcommand{\Bvec}{{\bm B}}

\newcommand{\es}[2] {\begin{equation} \label{#1} \begin{split} #2 \end{split} \end{equation}}

\begin{document}

\title{$X$-Ray Signatures of Axion Conversion in Magnetic White Dwarf Stars}
\author{Christopher Dessert}
\author{Andrew J. Long}
\author{Benjamin R. Safdi}
\affiliation{Leinweber Center for Theoretical Physics, University of Michigan, Ann Arbor, Michigan 48109, USA}

\date{\today}

\begin{abstract}
White dwarf (WD) stars may radiate keV-energy axions produced in their stellar cores.  This has been extensively studied as an extra channel by which WDs may cool, with some analyses even suggesting that axions can help explain the observed WD luminosity function.  We show that the radiated axions may convert into $X$-rays in the strong magnetic fields surrounding the WDs, leading to observable $X$-ray signatures.  We use {\it Suzaku} observations of the WD \mbox{RE J0317-853} to set the strongest constraints to date on the combination of the axion-electron ($g_{aee}$) times axion-photon ($g_{a\gamma\gamma}$) couplings, and we show that dedicated observations of magnetic WDs by telescopes such as {\it Chandra}, {\it XMM-Newton}, and {\it NuSTAR} could increase the sensitivity to $|g_{aee} g_{a\gamma\gamma}|$ by over an order of magnitude, allowing for a definitive test of the axion-like-particle explanation of the stellar cooling anomalies. 
\end{abstract}

\preprint{LCTP-19-05}

\maketitle

The quantum chromodynamics (QCD) axion, originally proposed to solve the strong {\it CP} problem~\cite{Peccei:1977ur,Peccei:1977hh,Weinberg:1977ma,Wilczek:1977pj},
is a well-motivated extension of the Standard Model of particle physics.  The QCD axion is a light pseudoscalar particle that couples to the QCD operator $G \tilde G$, with $G$ the QCD field strength.   
Additionally, the axion has dimension-5 couplings to electromagnetism and to matter.  
Studies of string compactifications show that, in addition to the QCD axion, there may exist a number of additional light pseudoscalar particles, with couplings to electromagnetism and matter but not to QCD~\cite{Svrcek:2006yi,Arvanitaki:2009fg}.
These pseudoscalars are called axion-like particles (ALPs), though throughout this work we will refer to all such particles as axions.  In this work we present a novel method, using $X$-ray observations of magnetic white dwarf (WD) stars (MWDs), to probe the existence of axions.

WDs have long been used as probes of axions by studying the possibility of energy loss from axion emission~\cite{Raffelt:1985nj}.  Axions are emitted by axion bremsstrahlung in electron-nucleon scattering.
The extra energy loss would modify WD cooling and thus change the luminosity function of WDs.  Comparisons to the observed luminosity function have been used to set stringent constraints on the axion-electron coupling~\cite{Isern:2008nt,Isern:2008fs,Bertolami:2014wua}. Moreover, it has been suggested that the observed WD luminosity function actually prefers the existence of an axion~\cite{Isern:2008nt,Isern:2008fs}, a claim further supported by period-drift measurements of WDs undergoing pulsations~\cite{Isern:2010wz}.

\begin{figure}[t]
\hspace{0pt}
\vspace{-0in}
\begin{center}
\includegraphics[width=0.45\textwidth]{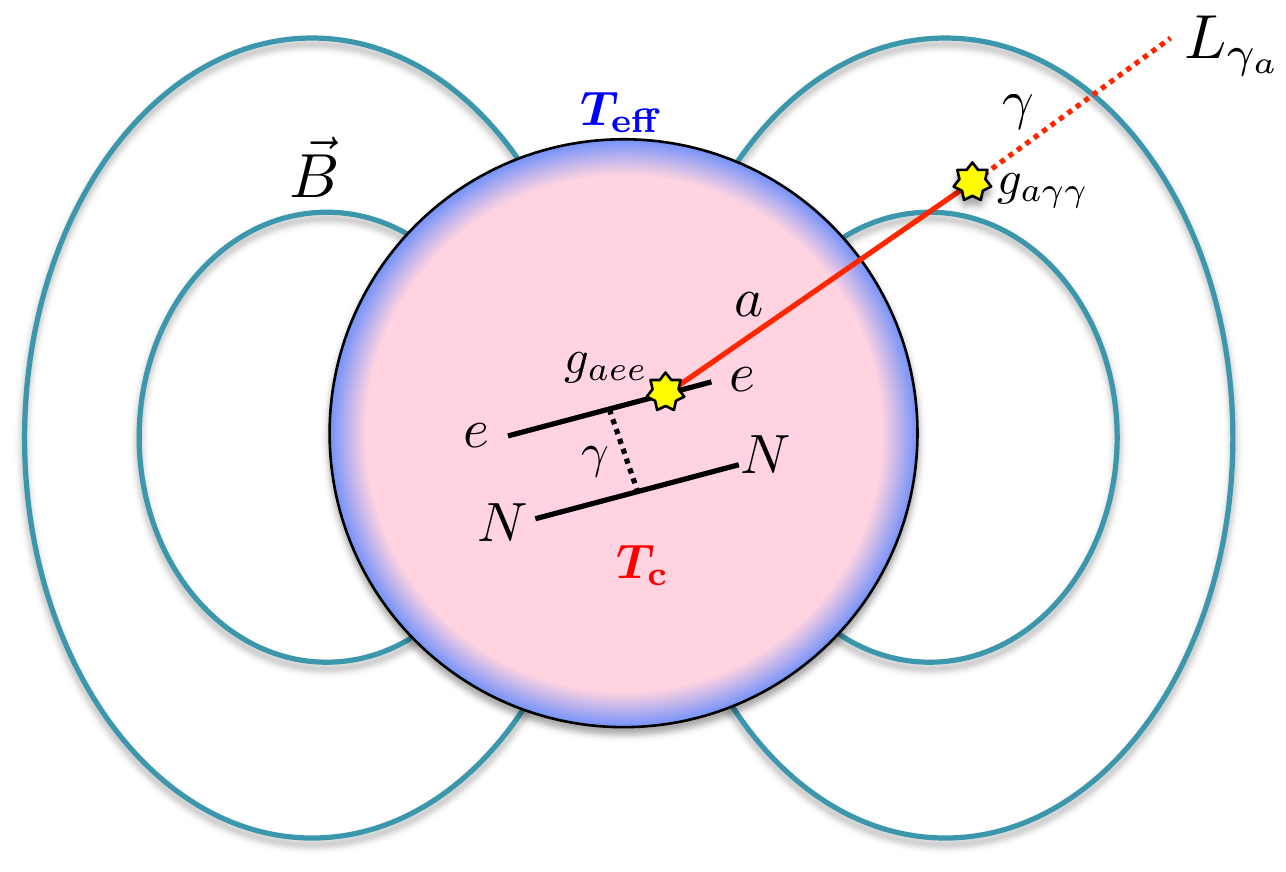}
\caption{
\label{fig:cartoon}
Axions are produced inside of a WD star and convert into $X$-ray photons as they pass through the star's magnetic field.  The axion-induced $X$-rays have energy around the core temperature $T_c$, which is much higher than the WD surface temperature $T_{\rm eff}$.   
}
\end{center}
\end{figure}

In this work we propose to use $X$-ray observations of MWDs to detect the small fraction of emitted axions that convert to $X$-ray photons outside of the MWD in the strong surrounding magnetic field, as illustrated in Fig.~\ref{fig:cartoon}.  The conversion of axions to photons utilizes the axion-photon coupling.  The proposed method uses the following key properties of isolated MWDs.  First, the surrounding magnetic fields can be quite high, $\sim$$10^9$ G~\cite{2013MNRAS.429.2934K,Ferrario:2015oda}, enhancing the axion-photon conversion probability.  Second, while the core temperature of WDs is typically in the $X$-ray band, $T_c \sim 10^7 \, \mathrm{K}$, the effective surface temperature is significantly lower, $T_{\rm eff} \sim 10^4 \, \mathrm{K}$ (see, {\it e.g.}, Ref.~\cite{Chabrier:2000ib}).  
Therefore an isolated MWD should not produce $X$-rays in the absence of axions.
  $X$-ray energy axions may escape the core and then convert into real $X$-ray photons in the magnetic field surrounding the WD, leading to a nearly thermal $X$-ray flux at the temperature of the core.  In this work, we show that $X$-ray telescope observations of MWDs have the potential to probe a wide range of axion masses and couplings.  However, as we will show, the $X$-ray observations are only sensitive to low-mass ALPs and not to the QCD axion, as the axion-photon conversion probability is suppressed for large axion masses. 

A similar approach to axion detection was previously suggested for neutron stars (NSs)~\cite{Morris:1984iz}.  In that case, nuclear bremsstrahlung in the NS generates an outgoing flux of axions, which may convert into $X$-rays in the strong magnetic field surrounding the NS.  However, it was pointed out in Ref.~\cite{Raffelt:1987im} that vacuum birefringence effects from quantum electrodynamics (QED), in the presence of strong magnetic fields, stymie the axion-photon conversion process.  The result is that the axion-induced $X$-ray flux from NSs is expected to be negligible for a KSVZ~\cite{Dine:1981rt,Zhitnitsky:1980tq,Srednicki:1985xd} or DFSZ~\cite{Kim:1979if,Shifman:1979if} QCD axion, but the flux may still be significant for ALPs~\cite{Fortin:2018ehg,Fortin:2018aom}.  We will discuss these QED effects for MWDs below.  

In many ways our proposal is reminiscent of the CAST experiment~\cite{Arik:2013nya,Arik:2015cjv,Anastassopoulos:2017ftl}, which looks for axions produced within the Sun.  These keV-energy axions travel to Earth where they are converted into $X$-ray photons by the strong magnetic field of the CAST experiment and detected with $X$-ray optics.  The key conceptual difference between our MWD proposal and CAST, in addition to using MWDs as the axion source instead of the Sun, is that the conversion to $X$-rays takes place not in the lab but rather in the magnetosphere surrounding the MWD itself.  A similar approach has in fact been suggested for the Sun (see, {\it e.g.}, Ref.~\cite{2009NJPh...11j5020Z}), whereby one looks at spectral and morphological distortions to the solar $X$-ray spectrum from axion-photon conversion in the solar magnetic field, but this is complicated by the fact that the Sun is already a strong $X$-ray source.  

We note that a host of additional astrophysical probes of axions and axion dark matter (DM) have been proposed.  These include radio signatures of axion DM conversion in NSs~\cite{Pshirkov:2007st,Huang:2018lxq,Hook:2018iia,Safdi:2018oeu} and radio signatures of axion decay~\cite{Caputo:2018vmy}, supernova cooling~\cite{Raffelt:2006cw,PhysRevD.94.085012}, energy loss in horizontal branch stars in globular clusters~\cite{PhysRevLett.113.191302}, and photon-axion oscillations leading to increased transparency of TeV gamma rays~\cite{Simet:2007sa,2009PhRvD..79l3511S,DeAngelis:2011id,TheFermi-LAT:2016zue}.
Axion DM is also the subject of significant laboratory efforts at present~\cite{Battaglieri:2017aum,Shokair:2014rna,Du:2018uak,Brubaker:2016ktl,Kenany:2016tta,Brubaker:2017rna,TheMADMAXWorkingGroup:2016hpc,Kahn:2016aff,Foster:2017hbq,Ouellet:2018beu,Chaudhuri:2014dla,Silva-Feaver:2016qhh,Budker:2013hfa,Bahre:2013ywa,Bogorad:2019pbu}.

{\quad} \\ 
\noindent
{\bf ${\bm X}$-ray flux calculation ---} 
It is useful to recall how an axion interacts with matter (see Ref.~\cite{Tanabashi:2018oca} for a review).  The QCD axion couples to gluons through the operator \mbox{${\mathcal L} \supset - a \alpha_s G \tilde G/(8 \pi f_a)$}, where $a$ is the axion field, $\alpha_s$ is the strong fine structure constant, $f_a$ is the axion decay constant, and $G$ is the QCD field strength tensor.  The axion may also couple to electromagnetism through the operator \mbox{${\mathcal L} \supset - g_{a\gamma\gamma} a F \tilde F/4$}, where $g_{a\gamma \gamma}$ is the axion-photon coupling with units of $\mathrm{GeV}^{-1}$, and $F$ is the QED field strength tensor.  Finally the axion can interact with electrons through the operator \mbox{${\mathcal L} \supset g_{aee} / (2 m_e) \bar e \gamma^\mu \gamma_5 e \partial_\mu a$}, where $g_{aee}$ is the dimensionless axion-electron coupling, $m_e$ is the electron's mass, and $e$ is the electron field.  It is customary to write \mbox{$g_{a\gamma \gamma} = C_\gamma \alpha_\EM / (2 \pi f_a)$} and \mbox{$g_{aee} = C_e m_e / f_a$}, where $\alpha_\EM$ is the electromagnetic fine structure constant, and $C_\gamma$ and $C_e$ are dimensionless parameters.

The coupling of axions to matter allows axions to be emitted from WDs. WDs have dense cores that are supported by electron degeneracy pressure and consist predominantly of a hot plasma of carbon, oxygen, and electrons.  The temperature of this isothermal plasma, which makes up the bulk of the matter of the WD, can be $T_c \approx 10^7$ K or higher, depending on the WD.  Axions are emitted most efficiently in WDs from bremsstrahlung off of electrons in electron-nuclei scattering, through the diagram illustrated in Fig.~\ref{fig:cartoon}. 
 The luminosity radiated from a WD in axions, denoted by $L_a$, is 
calculated in Refs.~\cite{Raffelt:1985nj,Nakagawa:1988rhp,Nakagawa:1987pga} (see also Ref.~\cite{Raffelt:1990yz}), and for a carbon-oxygen WD with mass density ratio $0.5$ and total mass $M_\WD$, we find
\es{eq:L_a}{
{L_a \over L_\odot} & \approx 1.6 \times 10^{-4}    \left({ g_{aee} \over 10^{-13}} \right)^2 \left( {M_\WD \over 1 \, M_\odot} \right)  \left( {T_c \over 10^7 \, \, \text{K}} \right)^4 \,,
}
where $L_\odot$ and $M_\odot$ are the luminosity and mass of the Sun, respectively.
The energy spectrum of the axion emission is found to be thermal at temperature $T_c$: \mbox{$dL_a/dE \propto E^3 / (e^{E/T_c} - 1)$}~\cite{Nakagawa:1988rhp}.

The emitted axions may be converted into $X$-ray photons in the magnetic field surrounding the MWD.  Under the approximation where we assume all axions travel along radial trajectories originating from the MWD center, we may derive a simple analytic expression for the conversion probability.  Note that in the Supplemental Material (SM) we derive and numerically solve the relevant equations of motion for the more general trajectories and field configurations appropriate for a realistic MWD with finite extent, but the results are similar to those found by assuming radial trajectories.  The axion-photon mixing equations in the presence of an external magnetic field can be reduced to a system of first-order differential equations using a WKB approximation.  This approximation assumes that the scale of variation in the magnetic field is much larger than the axion's de Broglie wavelength.  By working in the Weyl gauge, $A^0 = 0$, and 
focusing on axion trajectories along which the angle with respect to the magnetic field does not change, these equations take the form~\cite{Raffelt:1987im}
\es{eom}{
\left[ i \partial_r + E + 
\begin{pmatrix}
\Delta_{\parallel} & \Delta_B \\
\Delta_B & \Delta_a 
\end{pmatrix}
\right] 
\begin{pmatrix}
A_{\parallel} \\ a
\end{pmatrix}
= 0 \,,
}  
where $E$ is the axion's energy.  Here $A_\parallel(r)$ denotes the vector potential component in the plane normal to the direction of propagation and parallel to the external magnetic field, as a function of the radial coordinate $r$, while $a(r)$ is the axion field.  The probability for an axion to convert into a photon $p_{a\to\gamma}$ is determined by solving Eq. \eqref{eom} and comparing the magnitude squared of an initial pure axion state with the asymptotic solution for the electromagnetic vector potential.  In the background magnetic field, the axion-photon interaction induces a mixing, which is parametrized by $\Delta_B(r) = (g_{a\gamma\gamma}/2) B(r) \sin \Theta$, where $B(r)$ is the strength of the magnetic field at radius $r$ and $\Theta$ is the angle between the radial propagation direction and the magnetic field, which is $r$-independent.  The term $\Delta_a = - m_a^2 / (2 E)$ incorporates the axion mass and is responsible for the slightly different momenta between the axion and photon states.  Strong-field QED effects in vacuum give rise to the term $\Delta_{\parallel}(r) = (7/2) E \xi(r) \sin^2 \Theta$, with $\xi(r) = (\alpha_\EM / 45 \pi) [B(r) / B_{\rm crit}]^2$ with $B_{\rm crit} = m_e^2 / e \approx 4.41 \times 10^{13} \ \mathrm{G}$ the critical field strength.  In general $\Delta_{\parallel}$ also contains a term related to the photon's effective plasma mass $\omega_{\rm pl}$, $\Delta_{\parallel} = - \omega_\mathrm{pl}^2 / (2 E)$, but this term is subdominant to the QED one for the systems that we consider. 

Since the $B$ field strength decreases as we move farther away from the surface of the MWD, it is important to solve the equation of motion Eq.~\eqref{eom} including the changing magnetic field profile with $r$.
In the weak-mixing limit we can solve Eq.~\eqref{eom} using the formalism of time-dependent perturbation theory, and the axion-photon conversion probability is found to be~\cite{Raffelt:1987im,Fortin:2018aom}
\es{eq:pert-prob}{
p_{a \to \gamma} = \Bigl| \int_{R_\WD}^\infty \! \mathrm{d} r^\prime \Delta_B(r^\prime) \, e^{i \Delta_a r^\prime - i \int_{R_\WD}^{r^\prime} \! \mathrm{d} r^{\prime\prime} \Delta_{\parallel}(r^{\prime\prime})} \Bigr|^2 \,.
} 
The integral starts at the star's surface, $r = R_\WD$, since it is assumed that $X$-ray photons produced inside the MWD cannot escape.  To first approximation we may model the magnetic field outside of the MWD as a magnetic dipole, such that $B(r) = (R_\WD / r)^3 B_0$ with $B_0$ the field at the surface and $r > R_\WD$.  
 In the SM we consider more general magnetic field configurations, but the results are largely the same.  In general, we evaluate Eq.~\eqref{eq:pert-prob} numerically.  Note that for typical MWD parameters and asymptotically small $m_a$, the conversion probabilities are of order $10^{-4} \times (g_{a\gamma\gamma} / 10^{-11} \ \mathrm{GeV}^{-1})^2$. 

Having calculated both the spectrum of axion radiation, $dL_a / dE$, and the axion-to-photon conversion probability, $p_{a\to\gamma}$, the flux of axion-induced $X$-ray photons at Earth is calculated as 
\es{eq:dFdE_def}{
	\frac{dF_{\gamma_a}}{dE}(E) = \frac{dL_a}{dE}(E) \times p_{a \to \gamma}(E) \times \frac{1}{4 \pi d_\WD^2} \,,
}
where $d_\WD$ is the distance between Earth and the MWD.  Note that for a typical MWD $d_{\rm WD} \approx 30$ pc away, combining Eq.~\eqref{eq:dFdE_def} with the conversion probability estimate and the axion luminosity Eq.~\eqref{eq:L_a} leads to axion-induced $X$-ray fluxes of order $F_{\gamma a} \sim 10^{-15} \times (g_{aee} g_{a\gamma\gamma} / 10^{-24} \, \, {\rm GeV}^{-1})^2$ erg$/$cm$^2$$/$s at low axion masses, though in practice we compute $dF_{\gamma_a}/dE$ precisely using the above formalism given the properties of the MWD under consideration.  In practice we compute the integrated flux $F_{2-10}$, defined as the integral of Eq.~\eqref{eq:dFdE_def} from 2 to 10 keV, to compare to data.  

The only parameter that appears in Eq.\eqref{eq:dFdE_def} that cannot be measured directly for a given MWD is the core temperature, $T_c$, which affects the axion luminosity through Eq.\eqref{eq:L_a}.  The core temperature is not directly observable, since the thin WD atmosphere is largely opaque to radiation.  The effective temperature $T_{\rm eff}$ of the atmosphere, which determines the observed luminosity $L_{\gamma}$, is much smaller than the temperature of the isothermal core $T_c$.  Understanding the relation between $L_\gamma$ and $T_c$ requires detailed modeling of the layers connecting the atmosphere to the degenerate core (see {\it e.g.}, Refs.~\cite{Hansen:1999fa,Salaris:2000nm,Chabrier:2000ib,Renedo:2010vb}).  We use the result of models presented in Ref.~\cite{Chabrier:2000ib}, which finds that for luminosities above $L_\gamma \sim 10^{-3.8}$ $L_\odot$ and below $\sim$$10^{-1.5}$ $L_\odot$ the luminosity-core temperature relation is well approximated by
\es{eq:Tc_relation}{
T_c \simeq 
	\bigl( 3 \times 10^6 \, \, {\rm K} \bigr) \left( \frac{L_\gamma}{10^{-4} L_\odot} \right)^{0.4} \,.
}  
{As we discuss further in the SM, the WD models in Refs.~\cite{Hansen:1999fa,Salaris:2000nm,Chabrier:2000ib,Renedo:2010vb} agree to within $O(10\%)$ uncertainty.}  

It is important to note that while the effective temperatures $T_{\rm eff}$ are often at the level of $10^4$ K, the emission from the WD does not follow a pure thermal distribution at this temperature since higher frequencies probe deeper within the WD atmosphere~\cite{1976ApJ...206L..67S}.
  Still, for the hot WDs we consider, the expected thermal hard $X$-ray flux ({\it e.g.}, from 2 to 10 keV) is negligible compared to the axion-induced fluxes that will be probed.

{\quad} \\ 
\noindent
{\bf ${\bm X}$-ray data analysis and projections ---} Although there are over $70 000$ known WD stars within approximately $100 \, \mathrm{pc}$ of Earth~\cite{Brown:2018dum}, the total number of known MWDs with well-measured properties is only around $200$~\cite{Ferrario:2015oda}.  In the SM we list several MWDs that are expected to have the largest axion-induced $X$-ray flux, while in this section we focus on the most promising candidate.   

The magnetic WD \mbox{RE J0317-853}~\cite{1995MNRAS.277..971B} is especially hot, has a strong magnetic field, and is relatively nearby, making it an excellent target for $X$-ray searches for axions.  Optical and UV observations suggest an effective temperature between $T_{\rm eff} = 30 000$~K and $60 000$~K, while at the same time being incredibly massive, $M_\WD \sim 1.3$~$M_\odot$, and compact, $R_\WD \sim 0.003 - 0.004$~$R_\odot$~\cite{1995MNRAS.277..971B,2010A&A...524A..36K}.  Note that the best-fit temperature presented in Ref.~\cite{1995MNRAS.277..971B} is $T_{\rm eff} = 4.93_{-0.12}^{+0.22} \times 10^4$~K, though a wider range was considered in Ref.~\cite{2010A&A...524A..36K}. The MWD rotates with a period of $\sim$$725$ s, and over this period the surface magnetic field strength is seen to oscillate.  Reference~\cite{Burleigh:1998pqa} used time-resolved far-UV spectroscopic data from the {\it Hubble Space Telescope} to study the variations in the locations and magnitudes of Lyman $\alpha$ transition to measure the magnetic field profile.  They found that the observable surface magnetic field varies between 200 and 800 MG over the WD rotation period, which is well fit by an offset dipole model~\cite{Burleigh:1998pqa}. They also find a best-fit temperature $T_{\rm eff} \sim 40 000$~K.  

In the SM we model in detail the magnetic field structure of this MWD and assess the sensitivity of our results to uncertainties in the value of $T_\mathrm{eff}$, for example.  Here, however, we simply calculate the predicted $X$-ray flux using Eq.~\eqref{eq:pert-prob} with the most conservative choice of parameters for the magnetic field strength and effective temperature of \mbox{RE J0317-853}.  We use a central dipole field with surface-field value $B_0 = 200 \ \mathrm{MG}$ and $\sin\Theta = 1$, corresponding to the lowest surface-field value observed in Ref.~\cite{Burleigh:1998pqa} anywhere on the WD surface over its period. 
   We also take the fiducial values $T_{\rm eff} = 30 000$~K, $M_\WD = 1.32$~$M_\odot$, and $R_\WD = 0.00405$~$R_\odot$~\cite{2010A&A...524A..36K}, which let us infer $L_\gamma = 0.0120$ $L_\odot$ from the Stefan-Boltzmann law and $T_c = 2.0 \times 10^7$ K from Eq.~\eqref{eq:Tc_relation}.   We note that higher $T_{\rm eff}$ lead to higher flux rates, and our fiducial value for $T_{\rm eff}$ is the lowest end of the range discussed in Refs.~\cite{1995MNRAS.277..971B,2010A&A...524A..36K}.  Parallax measurements from {\it Gaia}-DR2~\cite{Brown:2018dum} place \mbox{RE J0317-853} at a distance of $d_\WD = 29.54 \pm 0.04$~pc from Earth.  

The axion-induced $X$-ray spectrum from \mbox{RE J0317-853} should peak around \mbox{$E \sim 3 \, T_c \sim 5 \, \text{keV}$}.  Observations with {\it Suzaku}, using approximately $60 \, \text{ks}$ of exposure time, detected no astrophysical $X$-ray emission from this MWD and set a flux limit in the $2$-$10 \, \text{keV}$ range of $F _{2-10} < 1.7 \times 10^{-13} \, \text{erg}/\text{cm}^2/\text{s}$ at 95\% confidence~\cite{doi:10.1093/pasj/65.4.73}.  We note that the limit in Ref.~\cite{doi:10.1093/pasj/65.4.73} required background subtraction and modeling; the limit itself is dominated by systematic uncertainties in modeling the cosmic $X$-ray background and the non-$X$-ray background.  We also caution that the limit in Ref.~\cite{doi:10.1093/pasj/65.4.73} is formally only valid for an energy spectrum that resembles the above backgrounds, given the energy-dependence of the {\it Suzaku} effective area, though we have checked that this only induces a $\sim$10\% difference in the predicted counts and thus can be ignored for our purposes.  We translate the flux limit into a 95\% constraint on the axion coupling constants $|g_{a\gamma\gamma} \, g_{aee}|$ using the fiducial values for \mbox{RE J0317-853} and the formalism for axion-photon conversion developed above.  Our results are presented in Fig.~\ref{fig:main-result}, which shows our constraint on the axion parameter space; the region above the blue curve is excluded at $95\%$ confidence for the fiducial stellar parameters ($T_\mathrm{eff}$, $B_0$, etc) given above.  Note that the nontrivial structure in the limit at high $m_a$ is due to transitioning across the regime where $\Delta_{a} \ll \Delta_\parallel$ and then again to the regime where $\Delta_a \gg R_\text{WD}$.   

\begin{figure}[t]
\hspace{0pt}
\vspace{-0.3in}
\begin{center}
\includegraphics[width=0.50\textwidth]{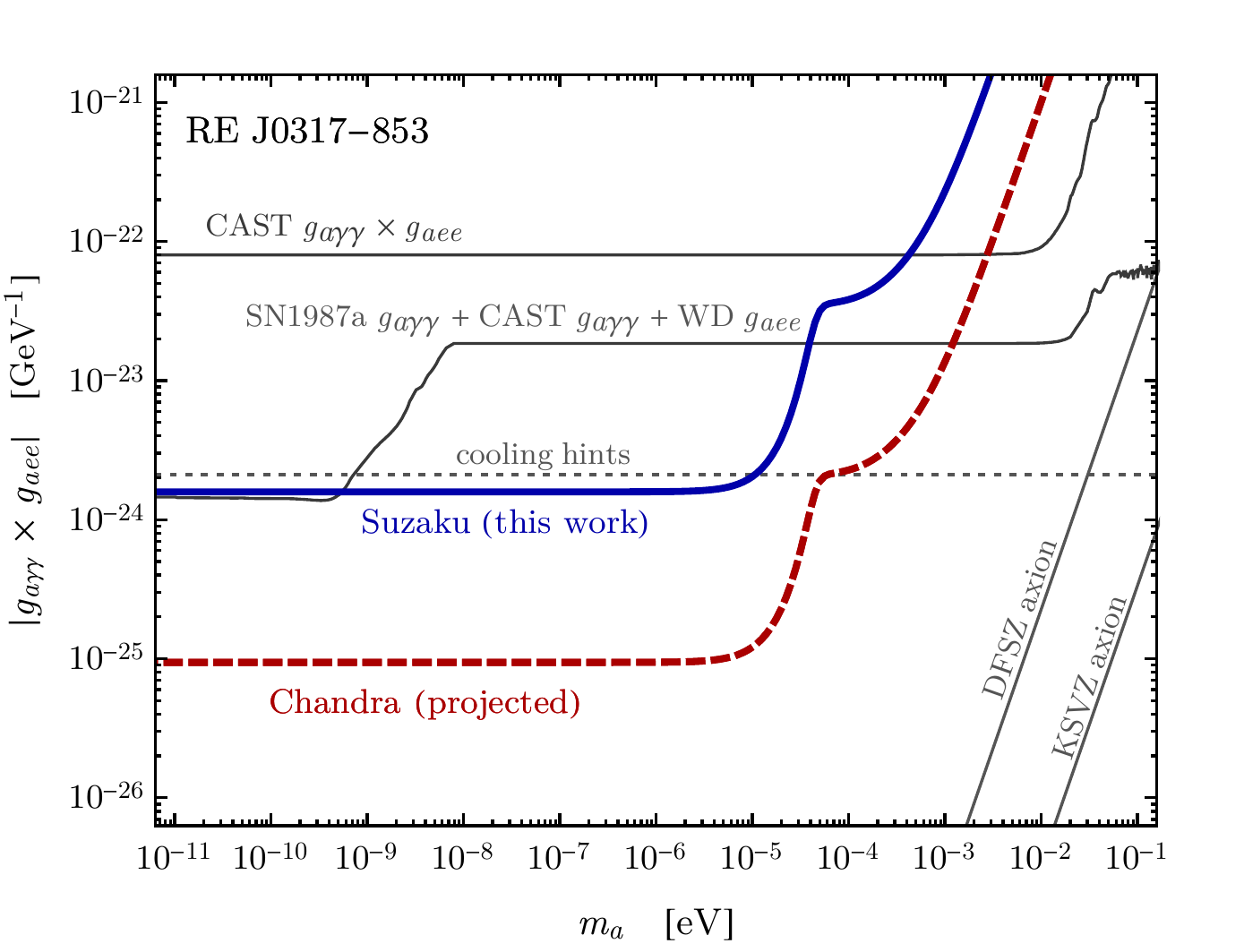}
\caption{
\label{fig:main-result}
The  $95\%$ C.L. upper limit on the axion couplings $|g_{a\gamma\gamma} \, g_{aee}|$ as a function of the axion mass $m_a$ are inferred from the nonobservation of $X$-ray emission from the MWD \mbox{RE J0317-853} by the {\it Suzaku} telescope with $\sim$$60 \, \mathrm{ks}$ of exposure (solid blue line).   
We also show the $95\%$ C.L. projected sensitivity from {\it Chandra} observations of the same MWD with a $400 \, \mathrm{ks}$ exposure (dashed red line). 
The limits extend to $m_a = 0$ outside of the plotted range.  The nontrivial structure in the limit at high $m_a$ arises from the transition probability becoming suppressed in this regime by the axion-photon momentum mismatch (see text for details). Additionally, we show the strongest upper limit on this parameter space before this work from the nonobservation of gamma rays from SN 1987A, searches for axions with the CAST experiment, and constraints on the WD luminosity function.  Stellar cooling hints suggest an axion may be present in the spectrum with $|g_{a\gamma\gamma} \, g_{aee}| \sim 2 \times 10^{-24}$, as indicated, though this interpretation is subject to large uncertainties.
}
\end{center}
\end{figure}

We compare our result to the previous best limits on this coupling combination.  By searching for axions produced inside of the Sun though the axion-electron coupling, the CAST experiment obtains a direct limit on $|g_{a\gamma\gamma} \, g_{aee}|$~\cite{Barth:2013sma}.  Our constraints are approximately 2 orders of magnitude stronger than those from CAST at low axion masses.
Several experiments derive upper limits on $|g_{a\gamma\gamma}|$ alone.  CAST provides an upper limit on $|g_{a\gamma\gamma}|$ from axions produced in the Sun through the Primakoff process~\cite{Anastassopoulos:2017ftl}.  Similarly an upper bound on $|g_{a\gamma\gamma}|$ is derived from the nonobservation of a gamma ray flux coincident with SN 1987A~\cite{Payez:2014xsa}, which is expected to be seen if axions are produced during the core-collapse supernova and subsequently convert into $\gamma$-ray photons in the intergalactic magnetic field.  By combining these limits on $g_{a\gamma\gamma}$ with the limit $|g_{aee}| < 2.8 \times 10^{-13}$, which arises from modeling the WD luminosity function~\cite{Bertolami:2014wua}, we obtain a second $95\%$ confidence upper limit, which is stronger than CAST's direct limit on $|g_{a\gamma\gamma} \, g_{aee}|$.  

In Fig.~\ref{fig:main-result} we also show the best-fit value for $g_{a\gamma\gamma} \times g_{aee}$ (``cooling hints'') found in Ref.~\cite{Giannotti:2017hny} from a global fit to the available stellar cooling data, which slightly favors nonzero axion-electron and axion-photon couplings.  Unfortunately the global fits say little about the axion mass, since the stellar cooling probes are not sensitive to the mass of the particle causing the cooling.  From our work we are able to say that the axion mass at these best-fit couplings needs to be above $\sim$$2\times 10^{-5}$ eV.  
The coupling-mass relations for the DFSZ~\cite{Kim:1979if,Shifman:1979if} and KSVZ~\cite{Dine:1981rt,Zhitnitsky:1980tq,Srednicki:1985xd} QCD axion models are also shown in Fig.~\ref{fig:main-result}. 

The {\it Suzaku} observations of MWD \mbox{RE J0317-853} from~\cite{doi:10.1093/pasj/65.4.73} are not ideal for searching for the proposed axion-induced $X$-ray signal.  This is because {\it Suzaku} has a rather poor point-source sensitivity in the $\sim$2-10 keV band compared to other telescopes like {\it XMM-Newton} or {\it Chandra}.  For instance, we estimate that 
 a $\sim$400 ks observation with {\it Chandra} 
would yield a 95\% confidence flux sensitivity in the $2$-$10$~keV band at the level of $6 \times 10^{-16} \, \text{erg}/\text{cm}^2/\text{s}$
~\cite{PIMMSweb}, 
which is over 2 orders of magnitude better than the sensitivity achieved in~\cite{doi:10.1093/pasj/65.4.73} with {\it Suzaku}. (Specifically, this flux sensitivity estimate includes $X$-ray and particle backgrounds as estimated in~\cite{PIMMSweb} and is for the ACIS-I instrument in the timed exposure mode with no grating and with CCD {\it I3}, along with Very Faint telemetry.)   This projected sensitivity appears in Fig.~\ref{fig:main-result} as the red dashed curve.  Note that this sensitivity projection assumes that the MWD does not produce $X$-ray emission in the energy range of interest at the flux levels that would be probed by {\it XMM-Newton} and {\it Chandra}.  This assumption is justified because the thermal emission is exponentially suppressed in the energy range 2-10 keV and well below the levels that would be probed by future observations.  Possible nonthermal emission mechanisms include synchrotron and curvature radiation in the strong magnetic field surrounding the MWD.  However, these processes are suppressed by the rather large spin period of the MWD, which leads to a small accelerating potential compared to {\it e.g.} the fast-spinning cataclysmic variable (CV) MWD AE Aquarii, for which pulsed nonthermal emission has been possibly observed~\cite{Terada:2007br,doi:10.1093/pasj/65.4.73}.  

{\quad} \\ 
\noindent
{\bf Discussion ---} 
In this Letter we used {\it Suzaku} observations of the nearby MWD \mbox{RE J0317-853} to set the strongest limits to date on $|g_{a\gamma\gamma} \, g_{aee}|$ for axion masses  $\lesssim$ meV.  Dedicated observations, with existing telescopes such as {\it XMM-Newton}, {\it Chandra}, or {\it NuSTAR} of this MWD and others have the potential to improve the sensitivity by over an order of magnitude and perhaps more.  If the previously observed stellar cooling hints are due to low-mass axion-like particles, $X$-ray observations of MWDs should detect excess hard $X$-ray flux.

If a hard $X$-ray signal is seen from a MWD, the first question should be if the MWD is accreting, which would be the case for a CV star.  CVs, which often emit hard $X$-ray spectra are distinguishable from {\it e.g.} their variability and emission line spectra.  It is also possible that some MWDs emit hard $X$-rays due to chromospheric activity, though this has yet to be observed~\cite{Brinkworth:2003xp}.  Our predicted axion signal has the unique feature of having an order unity modulation fraction over the MWD period for most alignment angle configurations, as discussed more in the SM, which could help differentiate it from other backgrounds.

The signal discussed in this work requires the axion to interact with both electromagnetism and electrons.  While both interactions are expected in a generic ALP theory, from an effective field theory point of view, this fact makes it hard to directly compare our sensitivity to {\it e.g.} the CAST limits on $g_{a\gamma\gamma}$ alone.  However, in the SM we show that even if one assumes that the axion-electron operator is generated through the renormalization group alone due to $W$ and $Z$-boson loops, the constraints from this work may be translated to constraints on $g_{a\gamma\gamma}$ that are comparable to those from CAST.  

In addition to the axion-electron coupling, a generic axion is also expected to interact derivatively with quarks.  These interactions cause NSs to radiate $\sim$keV axions, as in the WD case described in this work.
NSs are also promising targets for axion-induced hard $X$-ray signals, and this will be explored in future work~\cite{Buschmann:2019pfp}.  

\begin{acknowledgments}
{\it 
We are grateful to Raymond Co, Anson Hook, Yoni Kahn, and Nick Rodd for comments on the draft.  The work of C.D. and B.R.S. was supported in part by the DOE Early Career Grant No. DE-SC0019225.  A.J.L. was supported in part by the DOE under Grant No. DE-SC0007859. This research was supported in part through computational resources and services provided by Advanced Research Computing at the University of Michigan, Ann Arbor.  
}
\end{acknowledgments}

\bibliography{axion_from_WD}


\clearpage
\newpage
\maketitle
\onecolumngrid
\begin{center}
\textbf{\large $X$-ray signatures of axion conversion in magnetic white dwarf stars} \\ 
\vspace{0.05in}
{ \it \large Supplementary Material}\\ 
\vspace{0.05in}
{Christopher Dessert, \ Andrew J. Long, \ and \ Benjamin R. Safdi}
\end{center}
\onecolumngrid
\setcounter{equation}{0}
\setcounter{figure}{0}
\setcounter{table}{0}
\setcounter{section}{0}
\setcounter{page}{1}
\makeatletter
\renewcommand{\theequation}{S\arabic{equation}}
\renewcommand{\thefigure}{S\arabic{figure}}

This Supplementary Material is organized as follows.   Section~\ref{SM:candidates} includes a list of several MWD stars that are expected to be promising candidates for observations of axion-induced $X$-ray flux. In Sec.~\ref{SM:probability} we present a more general formalism for calculating the axion-to-photon conversion probability to account for the fact that the axions are emitted isotropically and homogeneously throughout the WD core.  In Sec.~\ref{SM:REJ0317-853} we perform a more detailed study of the $X$-ray emission from \mbox{RE J0317-853}, including a more accurate modeling of its magnetic field structure, an assessment of uncertainties in its temperature measurements, and an evaluation of its $X$-ray spectrum.  Finally Sec.~\ref{SM:gaee} presents the radiatively-induced axion-electron coupling that arises from the axion-photon coupling, which allows us to recast our limit on $|g_{a\gamma\gamma} \, g_{aee}|$ in terms of the axion-photon coupling alone.  

\section{Additional MWD candidates for $X$-ray observation}\label{SM:candidates}

In the main text we have focused our analysis on the MWD star \mbox{RE J0317-853}, since it is expected to have a particularly strong $X$-ray flux, and because $X$-ray data is already available from {\it Suzaku}.  However, there are over $200$ MWD stars with well-measured field strengths, temperatures, and distances.  For each of these stars, we calculate the expected axion-induced $X$-ray flux in the $2-10 \ \mathrm{keV}$ energy window, denoted by $F_{2-10}$, assuming $m_a = 10^{-9} \ \mathrm{eV}$ and $g_{a\gamma\gamma} g_{aee} = 10^{-24} \ \mathrm{GeV}^{-1}$.  The flux is insensitive to the axion mass in the limit $m_a \to 0$, and more generally the flux has an overall scaling with the couplings, $F_{2-10} \propto (g_{a\gamma\gamma} g_{aee})^2$.  

Our results are summarized in Table~\ref{tab:candidates}, which shows the ten MWDs with the largest predicted $X$-ray flux, $F_{2-10}$.  We constructed a full list by merging the SDSS DR7 magnetic WD catalog~\cite{Kleinman:2012nt}, a review MWD catalog~\cite{Ferrario:2015oda}, and Gaia DR2 WD catalog~\cite{Brown:2018dum}.  The former two provide the magnetic field strengths and temperatures of the WDs, while the latter provides distances, luminosities, masses, and radii, if known.  Since the mass and radius of \mbox{WD 2010+310} are not known, we take its mass to be $M_\WD = 1 \, M_\odot$, and we infer its radius from the Stefan-Boltzmann law.  

\begin{table}[h!]
\caption{\label{tab:candidates}MWD stars that make good candidates for measurement of their secondary, axion-induced $X$-ray flux.  The columns correspond to the star's mass in solar masses, radius in solar radii, luminosity in solar luminosities, effective temperature in Kelvin, magnetic field strength in mega-Gauss, distance from Earth in parsecs, and predicted $X$-ray flux from $2-10 \, \mathrm{keV}$ in $\mathrm{erg}/\mathrm{cm}^2/\mathrm{s}$, calculated assuming $m_a = 10^{-9} \, \mathrm{eV}$ and $g_{a\gamma\gamma} g_{aee} = 10^{-24} \ \mathrm{GeV}^{-1}$. The parameters were obtained by merging the catalogs in Refs.~\cite{Kleinman:2012nt,Ferrario:2015oda,Brown:2018dum}.  We infer the mass and radius of \mbox{WD 2010+310} as discussed in the text.  
}
\begin{ruledtabular}
\begin{tabular}{c|cllcccc}
& $M_\WD \ [M_\odot]$ & $R_\WD \ [R_\odot]$ & $L_\gamma \ [L_\odot]$ & $T_\mathrm{eff} \ [\mathrm{K}]$ & $B \ [\mathrm{MG}]$ & $d_\WD \ [\mathrm{pc}]$ & $F_{2-10} \ [\mathrm{erg}/\mathrm{cm}^2/\mathrm{s}]$ \\ \hline
\mbox{RE J0317-853} & $1.32$ & $0.00405$ & $0.0120$ & $30000$ & $200$ & $29.54$ & $6.8 \times 10^{-14}$ \\ 
\mbox{WD 2010+310} & $1^\ast$ & $0.00643^\ast$ & $0.00566$ & $19750$ & $520$ & $30.77$ & $4.4 \times 10^{-14}$ \\ 
\mbox{WD 0041-102 (Feige 7)} & $1.05$ & $0.00756$ & $0.00635$ & $18750$ & $35$ & $31.09$ & $3.0 \times 10^{-14}$ \\ 
\mbox{WD 1031+234} & $0.937$ & $0.00872$ & $0.0109$ & $20000$ & $200$ & $64.09$ & $2.3 \times 10^{-14}$ \\  
\mbox{WD 1533-057} & $0.717$ & $0.0114$ & $0.0121$ & $18000$ & $31$ & $68.96$ & $1.3 \times 10^{-14}$ \\  
\mbox{WD 1017+367} & $0.730$ & $0.0111$ & $0.0082$ & $16500$ & $65$ & $79.24$ & $7.1 \times 10^{-15}$ \\  
\mbox{WD 1043-050} & $1.02$ & $0.00787$ & $0.00388$ & $16250$ & $820$ & $83.33$ & $5.4 \times 10^{-15}$ \\  
\mbox{WD 1211-171} & $1.06$ & $0.00754$ & $0.00992$ & $21000$ & $50$ & $92.61$ & $5.4 \times 10^{-15}$ \\  
\mbox{SDSS 131508.97+093713.87} & $0.848$ & $0.00968$ & $0.01347$ & $20000$ & $14$ & $101.7$ & $3.5 \times 10^{-15}$ \\  
\mbox{WD 1743-520} & $1.13$ & $0.00681$ & $0.00184$ & $14500$ & $36$ & $38.93$ & $2.9 \times 10^{-15}$ \\  
\end{tabular}
\end{ruledtabular}
\end{table}

\section{The probability for axion-photon conversion in a general magnetic field background}\label{SM:probability}

In this section we present a more general formalism to calculate the axion-photon conversion probability, and in the following section we apply this calculation to study the $X$-ray emission from \mbox{RE J0317-853} in more detail.  Interactions between the axion and photon are described by the following Lagrangian: 
\begin{align}\label{eq:Lcal_aF}
	\mathcal{L} & = 
	\frac{1}{2} (\partial_\mu a)^2 
	- \frac{1}{2} m_a^2 a^2 
	- \frac{1}{4} F_{\mu\nu} F^{\mu\nu} 
	+ \frac{1}{2} m_A^2 A_\mu A^\mu 
	\\ & \qquad 
	- A_\mu j^\mu 
	- \frac{1}{4} g_{a\gamma\gamma} \, a \, F_{\mu\nu} \tilde{F}^{\mu\nu} 
	+ \frac{\alpha_\EM^2}{90 \, m_e^4} \Bigl[ \bigl( F_{\mu\nu} F^{\mu\nu} \bigr)^2 + \frac{7}{4} \bigl( F_{\mu\nu} \tilde{F}^{\mu\nu} \bigr)^2 \Bigr] \,,
	\nonumber
\end{align}
where $m_a$ is the axion's mass, $m_A$ is the photon's effective mass, $j^\mu$ is the electromagnetic current density, $g_{a\gamma\gamma}$ is the axion-photon coupling, $\alpha_\EM$ is the electromagnetic fine structure constant, and $m_e$ is the electron's mass.  An environment with a nonzero electron density, $n_e$, gives rise to an effective mass, $m_A = \omega_\mathrm{pl} = \sqrt{4 \pi \alpha_\EM n_e / m_e}$, and although this term will be negligible for our calculation, we retain it here for the sake of generality.  The last term in \eqref{eq:Lcal_aF}, known as the Euler-Heisenberg term, describes the photon's self-interaction that arises at energies below the electron's mass, where the electron can be integrated out of the theory.  

One can derive the field equations for $a({\bm x},t)$ and $A_\mu({\bm x},t)$ by applying the variational principle to \eqref{eq:Lcal_aF}.  It is customary to work in the Weyl gauge, $A_0 = A^0 = 0$~\cite{Raffelt:1987im}.  Upon writing $A_i = - A^i = - \Avec_i$ and $j^\mu = (\rho, {\bm j})$, the field equations are expressed as 
\begin{subequations}\label{eq:field_eqns_Weyl}
\begin{align}
	\ddot{a} 
	- \nabla^2 a 
	+ m_a^2 a 
	& \ = \ 
	- g_{a\gamma\gamma} \, \dot{\Avec} \cdot ({\bm \nabla} \times \Avec) \\ 
	- {\bm \nabla} \cdot \dot{\Avec} 
	& \ = \ 
	\rho 
	- g_{a\gamma\gamma} \, {\bm \nabla} a \cdot ({\bm \nabla} \times \Avec) 
	\\ & \qquad 
	+ \frac{16 \alpha_\EM^2}{45 \, m_e^4} {\bm \nabla} \cdot \Bigl[ 
	\frac{1}{2} \bigl( |\dot{\Avec}|^2 - |{\bm \nabla} \times \Avec|^2 \bigr) \dot{\Avec} 
	+ \frac{7}{4} \dot{\Avec} \cdot ({\bm \nabla} \times \Avec) ({\bm \nabla} \times \Avec) 
	\Bigr] \nonumber \\  
	\ddot{\Avec} 
	- \nabla^2 \Avec 
	+ {\bm \nabla} ({\bm \nabla} \cdot \Avec) 
	+ m_A^2 \Avec 
	& \ = \ 
	{\bm j} 
	+ g_{a\gamma\gamma} \, \dot{a} \, {\bm \nabla} \times \Avec 
	- g_{a\gamma\gamma} \, {\bm \nabla} a \times \dot{\Avec} 
	\\ & \qquad 
	- \frac{16\alpha_\EM^2}{45 \, m_e^4} \partial_t \Bigl[ 
	\frac{1}{2} \bigl( |\dot{\Avec}|^2 - |{\bm \nabla} \times \Avec|^2 \bigr) \dot{\Avec} 
	+ \frac{7}{4} \dot{\Avec} \cdot ({\bm \nabla} \times \Avec) ({\bm \nabla} \times \Avec) 
	\Bigr] 
	\nonumber \\ & \qquad 
	- \frac{16\alpha_\EM^2}{45 \, m_e^4} {\bm \nabla} \times \Bigl[ 
	\frac{1}{2} \bigl( |\dot{\Avec}|^2 - |{\bm \nabla} \times \Avec|^2 \bigr) ({\bm \nabla} \times \Avec) 
	- \frac{7}{4} \dot{\Avec} \cdot ({\bm \nabla} \times \Avec) \dot{\Avec} 
	\Bigr] \nonumber
	\, . 
\end{align}
\end{subequations}
Here $\partial_0 = \partial_t$ is denoted by a dot, and $\partial_i = \dvec_i$.  

To study axion-photon conversion, we are interested in the field equations for linearized perturbations around static background fields.  This motivates us to write 
\begin{align}\label{eq:bkg_expand}
	\rho({\bm x},t) = \bar{\rho}({\bm x}) 
	\ , \quad 
	{\bm j}({\bm x},t) = \bar{\bm j}({\bm x}) 
	\ , \quad 
	a({\bm x},t) = \bar{a}({\bm x}) + \delta a({\bm x},t) 
	\ , \quad \text{and} \quad 
	\Avec({\bm x},t) = \bar{\Avec}({\bm x}) + \dAvec({\bm x},t) \,,
\end{align}
where $\bar{a}({\bm x})$ and $\bar\Bvec({\bm x}) = \dvec \times \bar\Avec$ are the background axion and magnetic fields.  In particular, we are interested in systems for which $\bar{a} = 0$ and $\bar\Bvec$ can be approximated as a magnetic dipole.  

We assume that gradients in the background field, $\bar\Bvec$, are small compared to the wavenumber of the axions, $k = |{\bm k}|$, which lets us apply the WKB approximation.  Consider the trajectory ${\bm x}(z) = {\bm x}_0 + z \, \hat{\bm k}$ where ${\bm x}_0$ is a point inside the WD, $z \geq 0$ parametrizes the distance, and the unit vector $\hat{\bm k} = {\bm k} / k$ is the trajectory's orientation.  
If gradients in the directions transverse to $\hat{{\bm k}}$ are small, then we are motivated to adopt the plane-wave Ansatz
\begin{subequations}
\begin{align}
	\delta a({\bm x},t) & = a(z) + \mathrm{c.c.} \\
	& \text{with} \qquad a(z) = \tilde{a}(z) \, e^{-i \omega t + i k z} \nonumber \\ 
	\dAvec({\bm x},t) & = -i \bigl[ A_x(z) \, \hat{\bm e}_x + A_y(z) \, \hat{\bm e}_y + A_z(z) \, \hat{\bm e}_z \bigr] + \, \mathrm{c.c.} \\ 
	& \text{with} \qquad A_a(z) = \tilde{A}_a(z) \, e^{-i \omega t + i k z} \nonumber
	\, .
\end{align}
\end{subequations}
For relativistic particles, it is an excellent approximation to take the natural frequency as $\omega \approx k$.  
We choose a set of orthonormal basis vectors, $\{ \hat{\bm e}_x, \, \hat{\bm e}_y, \, \hat{\bm e}_z \}$, such that $\hat{\bm e}_z = \hat{\bm k}$ and therefore $\hat{\bm e}_x \cdot \hat{\bm k} = \hat{\bm e}_y \cdot \hat{\bm k} = 0$.  
We can also write $B_x(z) = \bar\Bvec \cdot {\bm e}_x$, $B_y(z) = \bar\Bvec \cdot {\bm e}_y$, $B_T(z) = [B_x^2 + B_y^2]^{1/2}$, $B_L(z) = \bar\Bvec \cdot \hat{\bm k}$, and $B(z) = |\bar\Bvec| = [B_T^2 + B_L^2]^{1/2}$.  
In the WKB approximation, we assume that the background changes slowly along the axion's trajectory.  This lets us drop derivatives of the background field and also second derivatives of the perturbations.  With these simplifications, the Weyl-gauge field equations~\eqref{eq:field_eqns_Weyl} become 
\begin{align}\label{eq:matrix_eqn}
	\left[ i \partial_z + k + \begin{pmatrix} \Delta_x & \Delta_{xy} & \Delta_{ax} \\ \Delta_{xy} & \Delta_y & \Delta_{ay} \\ \Delta_{ax} & \Delta_{ay} & \Delta_a \end{pmatrix} \right] \begin{pmatrix} A_x(z) \\ A_y(z) \\ a(z) \end{pmatrix} = 0 \,,
\end{align}
where we have defined 
\begin{subequations}
\begin{align}
	\Delta_a & \equiv \frac{1}{2k} \bigl( \omega^2 - k^2 - m_a^2 \bigr) \\ 
	\Delta_x & \equiv \frac{1}{2k} \bigl( \omega^2 - k^2 - m_A^2 \bigr) + \frac{8\alpha_\EM^2}{45 \, m_e^4} \Bigl( \frac{7}{4} \frac{\omega^2}{k} B_x^2 + k B_y^2 \Bigr) \\ 
	\Delta_y & \equiv \frac{1}{2k} \bigl( \omega^2 - k^2 - m_A^2 \bigr) + \frac{8\alpha_\EM^2}{45 \, m_e^4} \Bigl( \frac{7}{4} \frac{\omega^2}{k} B_y^2 + k B_x^2 \Bigr) \\ 
	\Delta_{ax} & \equiv \frac{1}{2} g_{a\gamma\gamma} \frac{\omega}{k} B_x \\ 
	\Delta_{ay} & \equiv \frac{1}{2} g_{a\gamma\gamma} \frac{\omega}{k} B_y \\ 
	\Delta_{xy} & \equiv \frac{8\alpha_\EM^2}{45 \, m_e^4} \Bigl( \frac{7}{4} \frac{\omega^2}{k} - k \Bigr) B_x B_y 
	\, .
\end{align}
\end{subequations}
In deriving \eqref{eq:matrix_eqn}, we have dropped terms that are nonlinear in the perturbations.  The wavefunction $\tilde{A}_z$ is not dynamical, meaning that its field equation is algebraic rather than differential, and we can remove it from the system of equations.  
We regain \eqref{eom} in the main text by taking $z=r$, $\omega = k$, $m_A = 0$, and by focusing on isotropic field configurations for which $B_x = 0$, $B_y(r) = B(r) \sin \Theta$, and $B_z(r) = B(r) \cos \Theta$.  

The axion-to-photon conversion probability is calculated by solving \eqref{eq:matrix_eqn}.  We assume that photons produced inside of the star, $r < R_\WD$, will be unable to escape as $X$-ray emission, and therefore we are only interested in conversions that occur outside of the star.  Along a given trajectory, let $z = z_0$ denote the surface of the star.  We solve \eqref{eq:matrix_eqn} along with the initial condition $a(z_0) = a_0$, $A_x(z_0) = 0$, and $A_y(z_0) = 0$, and then the axion-to-photon conversion probability is given by $p_{a\to\gamma} = \bigl( |A_x(\infty)|^2 + |A_y(\infty)|^2 \bigr) / a_0^2$.  Since \eqref{eq:matrix_eqn} is linear, the initial condition $a_0$ cancels when calculating $p_{a\to\gamma}$.  

\section{Analytic approximations to the conversion probability}\label{SM:prob_approx}

In order to develop our intuition, let us 
consider axion-photon conversion in a constant magnetic field, $B(r) = B$, which extends for a distance $\Delta r = R_\WD$ outside of the MWD, with $R_\WD$ the star's radius.  Then~\eqref{eom} may be solved exactly, and the axion-photon conversion probability is found to be 
\es{eq:p_a_to_gam}{
	p_{a\to\gamma}  = \sin^2 2 \theta \ \sin^2 (\Delta_\mathrm{osc} R_\WD/2) \,,
}
where the mixing angle, $\theta$, and the oscillation length scale, $\Delta_\mathrm{osc}^{-1}$, can be written as 
\es{}{
	\tan 2 \theta & = \frac{2 \Delta_B}{\Delta_{\parallel} - \Delta_a} \,, \qquad \Delta_\mathrm{osc} = {\Delta_{\parallel} - \Delta_a \over \cos 2 \theta} \,.
}
Note that $\Delta_{\parallel} - \Delta_a$ is always positive.  

We will generally be interested in the weak-mixing regime where $\theta \ll 1$, and for the sake of illustration let us also focus on the regime where the axion mass is low, such that $|\Delta_a| \ll \Delta_\parallel$.  With these assumptions, \eqref{eq:p_a_to_gam} becomes $p_{a\to\gamma} \approx 4 (\Delta_B^2 / \Delta_\parallel^2) \, \sin^2 (\Delta_\parallel R_\WD/2)$.  If the magnetic field is weak, such that $\Delta_{\parallel} R_\WD \ll 1$, then the conversion probability becomes $p_{a\to\gamma} \approx \Delta_B^2 R_\WD^2 \propto g_{a\gamma\gamma}^2 B^2$.  Alternatively, if the magnetic field is very strong, such that $1 \ll \Delta_\parallel R_\WD$, then we have instead $p_{a\to \gamma} \propto g_{a\gamma\gamma}^2 / (B E)^2$.  Here we see that the conversion probability is suppressed in the strong-magnetic field regime, which is a consequence of the QED birefringence effects, as anticipated in~\cite{Raffelt:1987im}.  The axion-photon conversion probability is maximal for $\Delta_\mathrm{osc} R_\WD \sim 1$.  In practice, for frequencies $E \sim 1$ (10) keV, this transition occurs for magnetic fields around $\sim$$10^7$ ($\sim$$5 \times 10^6$) G.  The dipole magnetic fields around the MWDs fall with distance from the WD surface, which leads to the result that if the magnetic field at the surface of the WD is high, the conversion probability is suppressed until the field drops to where $\Delta_\mathrm{osc} R_\WD \sim 1$. 
  
As we increase the axion mass, we enter the regime where $\Delta_\parallel \ll |\Delta_a|$ and $p_{a\to\gamma} \approx 4 (\Delta_B^2 / \Delta_a^2) \, \sin^2 (\Delta_a R_\WD/2)$.  For asymptotically large axion masses we have also $|\Delta_a| R_\WD \gg 1$, in which case $p_{a\to\gamma} \propto g_{a\gamma\gamma}^2 B^2 E^2 / m_a^4$.  Thus, contrary to the search for low-mass axions, searches for high-mass axions benefit from the largest magnetic fields and largest frequencies available.  

With the dipole field configuration we instead need to compute the integral given in~\eqref{eq:pert-prob}.  While in general this must be done numerically, we note that an analytic approximation is available in the regime of small axion mass, such that $|\Delta_a| \ll \Delta_\parallel$, where we find 
\es{eq:prob_approx}{
	p_{a \to \gamma} & \approx \frac{(\Delta_{B,0} R_\WD)^2}{(\Delta_{\parallel,0} R_\WD)^{\frac{4}{5}}}
	\Bigl| \frac{\Gamma(\frac{2}{5}) - \Gamma(\frac{2}{5}, -\frac{i}{5} \, \Delta_{\parallel,0} R_\WD) \bigr]}{5^{\frac{3}{5}}} \Bigr|^2 \,,
}
with $\Gamma(z)$ the gamma function, $\Gamma(a,z)$ is the incomplete gamma function, $\Delta_{B,0} = \Delta_B(R_\WD)$, and $\Delta_{\parallel,0} = \Delta_\parallel(R_\WD)$.  

\begin{figure}[t]
\hspace{0pt}
\vspace{-0in}
\begin{center}
\includegraphics[width=0.45\textwidth]{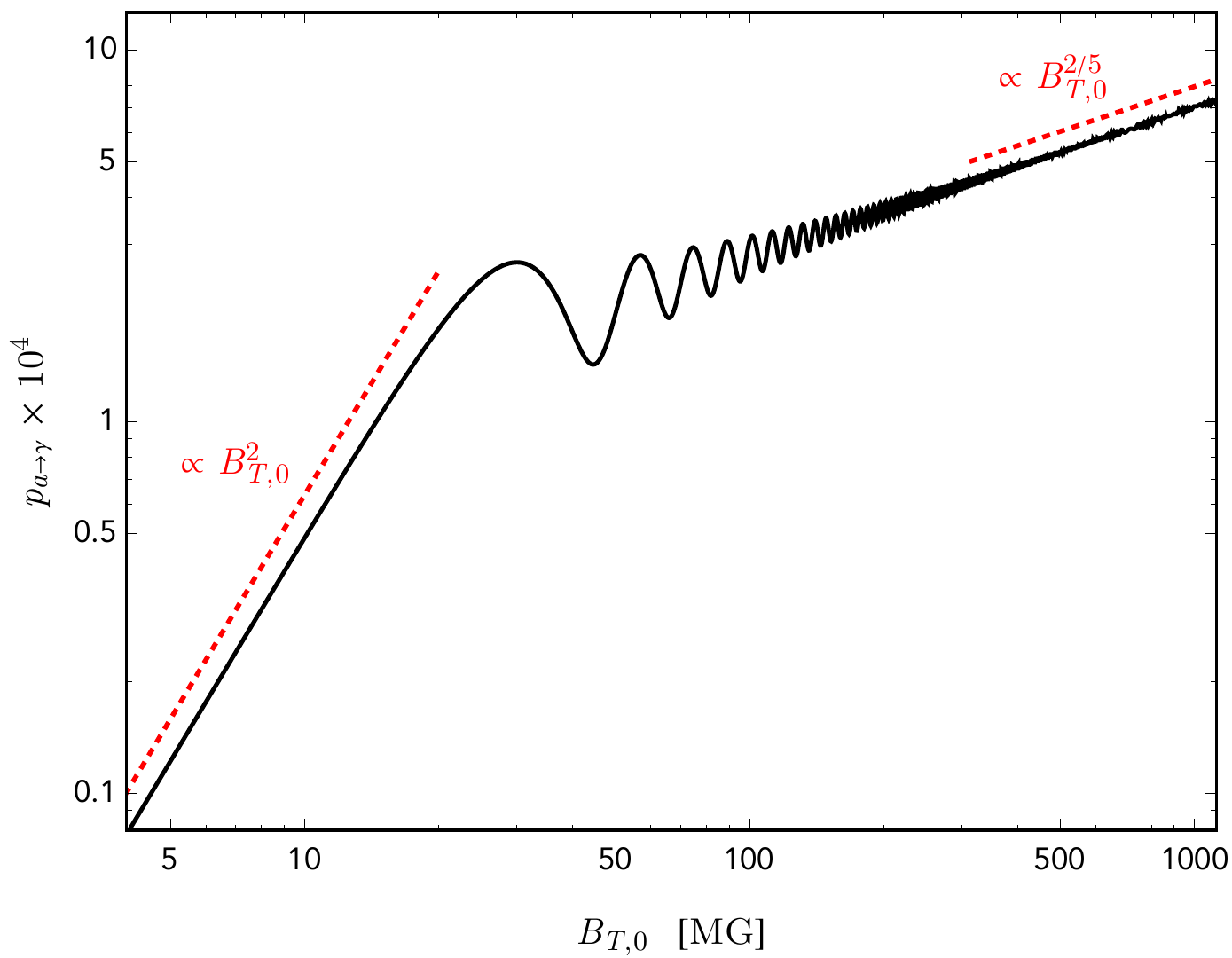}
\caption{
\label{fig:prob_versus_B}
The probability $p_{a\to\gamma}$ for axion to convert into $X$-ray photons in the presence of a dipole magnetic field with surface transverse field strength $B_{T,0}$.  We use~\eqref{eq:prob_approx} with $m_a = 0$, $g_{a\gamma\gamma} = 10^{-11} \ \mathrm{GeV}^{-1}$, $R_\WD = 0.00405$~$R_\odot$, and $E = 10 \ \mathrm{keV}$.  In general $p_{a\to\gamma} \propto g_{a\gamma\gamma}^2$.  
}
\end{center}
\end{figure}

To help illustrate the different regimes for the conversion probability, in Fig.~\ref{fig:prob_versus_B} we illustrate the conversion probability computed using~\eqref{eq:prob_approx} as a function of the transverse magnetic field $B_{T,0}$ at $r = r_\WD$ for an axion with an arbitrary small mass, a reference $g_{a\gamma\gamma} = 10^{-11} \ \mathrm{GeV}^{-1}$, and a MWD with $R_\WD = 0.00405$~$R_\odot$.  For definiteness we show the conversion probability for $E = 10 \ \mathrm{keV}$.  At small values of $B_{T,0}$ the conversion probability rises like $B_{T,0}^2$.  However, the Euler-Heisenberg term becomes important for $B_{T,0} \gtrsim 10$ MG, causing the conversion probability to scale less strongly with $B_{T,0}$ beyond this point. 

\section{A detailed analysis of the axion-induced $X$-ray flux from \mbox{RE J0317-853}}\label{SM:REJ0317-853}

Our analysis of the $X$-ray flux from \mbox{RE J0317-853} in the main text makes several simplifying assumptions.  In this section we revisit and refine these assumptions.  

\subsection{Magnetic field structure}  
In the main text we calculated the probability for axions to convert into $X$-ray photons in a background magnetic field by assuming that the background field is isotropic along the axion's trajectory and decreases with propagation distance like $B \sim r^{-3}$.  This is a reasonable assumption for a dipole field configuration if axions propagate on radial trajectories, as if they were produced at the star's center.  However, in general the axion's trajectory will not originate at the origin and, moreover, the background field will deviate from a simple dipole configuration.  Here we assess the effect of relaxing these assumptions.  

\begin{figure}[t]
\hspace{0pt}
\vspace{-0in}
\begin{center}
\includegraphics[width=0.45\textwidth]{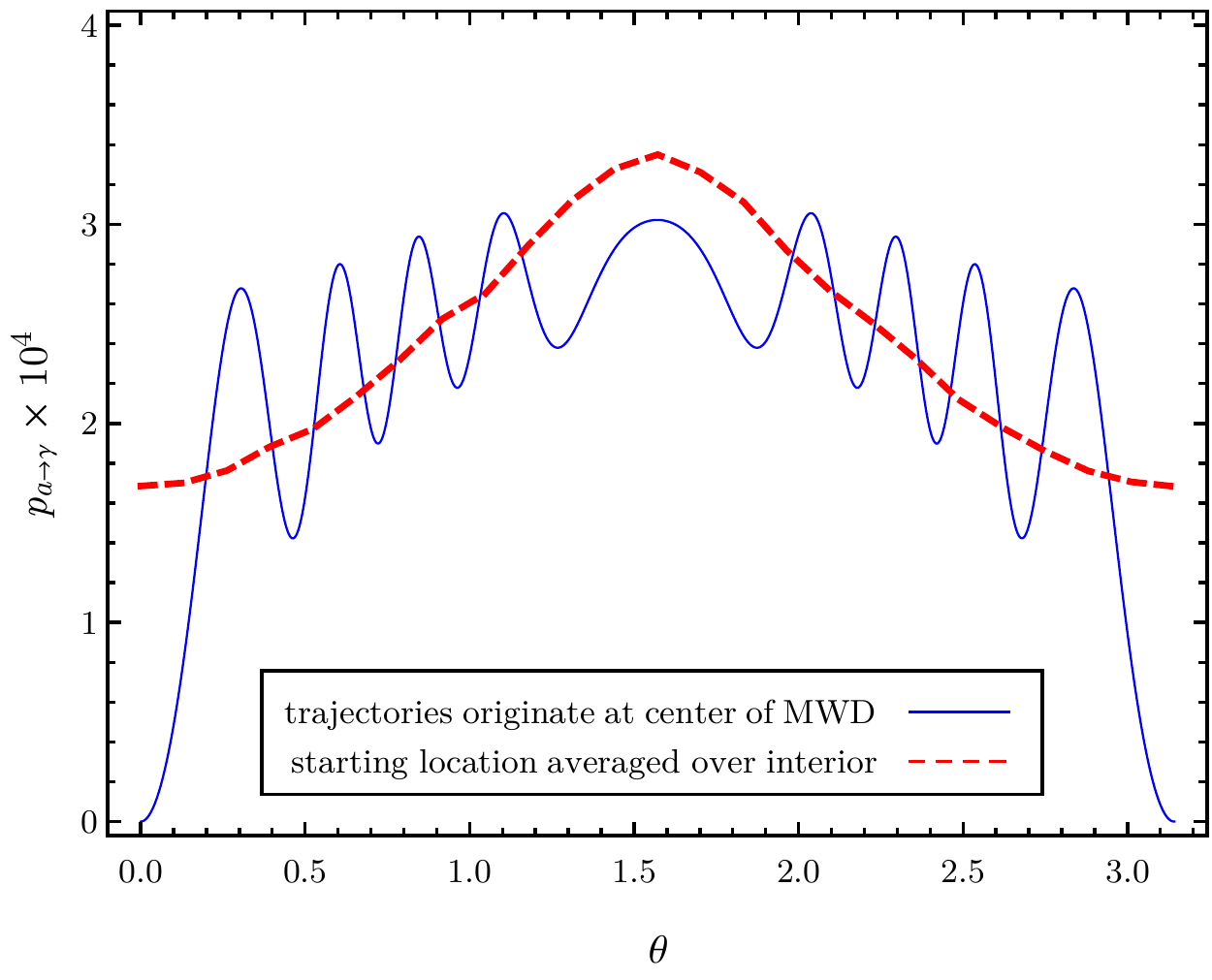} \hfill
\includegraphics[width=0.45\textwidth]{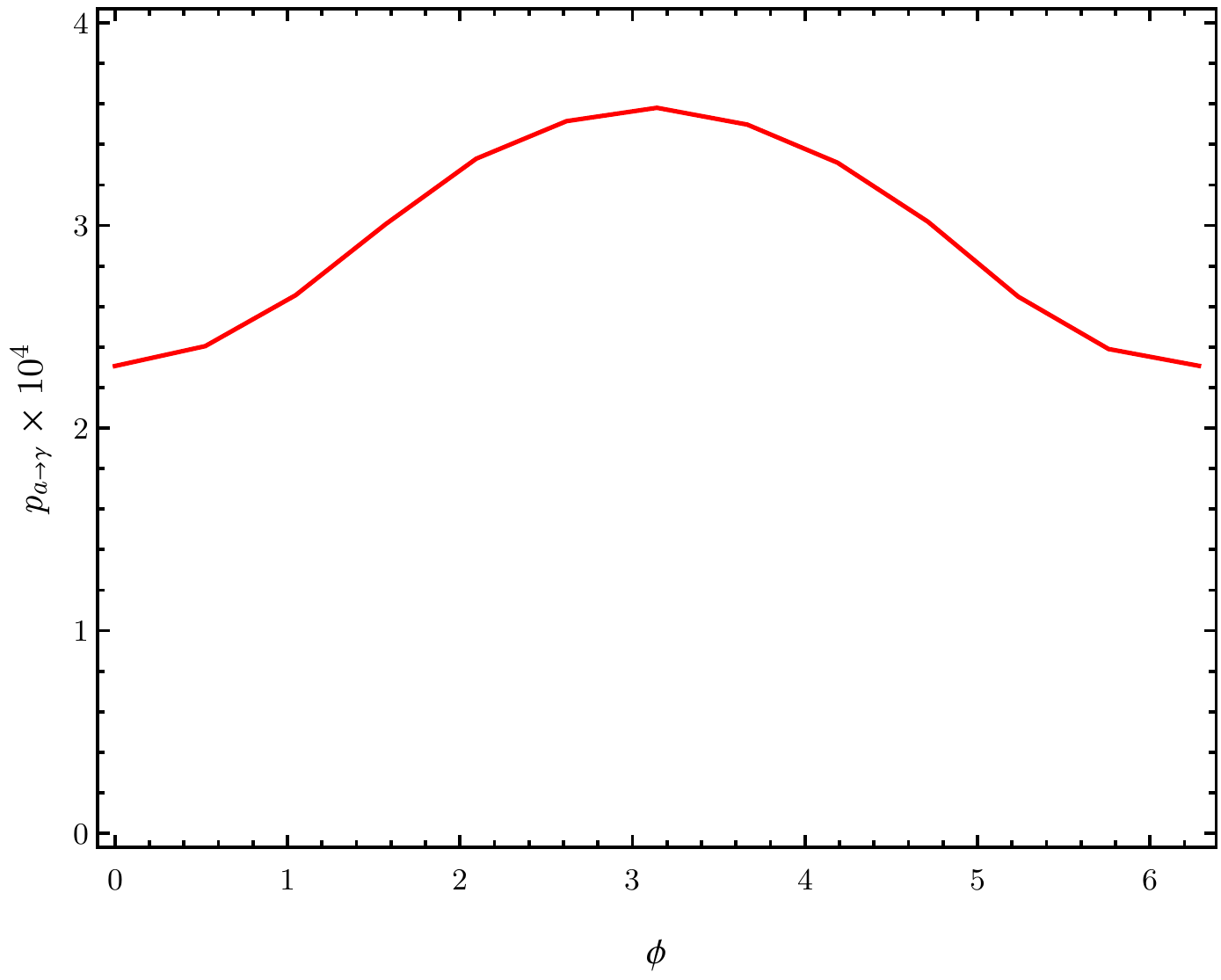}
\caption{
\label{fig:prob_theta}
The axion-photon conversion probability, $p_{a\to\gamma}$, for different models of the magnetic field around \mbox{RE J0317-853}.  
{\it Left:}  We assume a magnetic dipole field with polar field strength $B_0 = 400 \, \mathrm{MG}$ (200 MG at $\theta = \pi/2$), and we calculate $p_{a\to\gamma}$ for axion trajectories that propagate radially outward from the star's center (blue) as a function of the angle $\theta$ between the magnetic pole and the propagation direction.  We also show an average over trajectories that originate throughout the star's interior (red, dashed).  
{\it Right:}  We assume the offset-dipole model of~\cite{2010A&A...524A..36K} and calculate the trajectory-averaged conversion probability as a function of the azimuthal angle $\phi$ measured from the star's rotation axis, which corresponds to the phase over the $\sim$$725$ s period.  
For both panels, we have taken $m_a = 10^{-9} \, \mathrm{eV}$, $g_{a\gamma\gamma} = 10^{-11} \, \mathrm{GeV}^{-1}$, $\omega = 10 \, \mathrm{keV}$, and $R_\WD = 0.00405 \, R_\odot$.  
}
\end{center}
\end{figure}

Let us first understand how our results depend on the axion's trajectory through the magnetosphere.  Consider a dipolar magnetic field centered on the center of the star.  Such a field can be written as 
\begin{align}\label{eq:B_dipole}
	\bar\Bvec = \frac{|{\bm m}|}{4\pi} \, \frac{1}{r^3} \, \Bigl( 3 (\hat{\bm m} \cdot \hat{\bm x}) \, \hat{\bm x} - \hat{\bm m} \Bigr) 
	\qquad \text{for} \qquad r > R_\WD \,,
\end{align}
where ${\bm m} = |{\bm m}| \, \hat{\bm m}$ is the magnetic dipole moment, $|{\bm m}| / 4\pi = B_0 R_\WD^3 / 2$ is the relationship with the polar field $B_0$, ${\bm x} = r \, \hat{\bm x}$ is the spatial coordinate, and $r = |{\bm x}|$ is the distance from the center of the MWD.  Note that here $B_0$ is the value of the magnetic field at the star's surface in the direction of the magnetic pole, whereas in the main text below \eqref{eq:pert-prob} we used $B_0$ to denote the value at the surface in any arbitrary direction.
Note that the orientation of $\bar{\bm B}$ along a radial trajectory is invariant, which allows us to calculate the axion-photon conversion probability, $p_{a\to\gamma}$, using \eqref{eq:pert-prob}, in this case.  The result is presented in the left panel of Fig.~\ref{fig:prob_theta} as the blue curve, where we show the conversion probability as a function of the angle between the magnetic pole and the viewing angle, $\hat{\bm m} \cdot \hat{\bm x} = \cos \theta$.  As one expects, the probability vanishes at $\theta = 0$ and $\pi$ where the transverse magnetic field vanishes, and the probability peaks at $\theta = \pi/2$ where the transverse magnetic field, $B_T = B \, \sin \theta$, is maximal.  It is a rough approximation to suppose that axions only propagate on radial trajectories, and more accurately we should allow the axion's trajectory to originate anywhere inside the MWD.  In general the field direction is not invariant along such a trajectory, and we cannot use \eqref{eq:pert-prob} to calculate $p_{a\to\gamma}$, but rather we must solve \eqref{eq:matrix_eqn} directly.  Doing so leads to the red-dashed curve in the left panel of Fig.~\ref{fig:prob_theta}, where we have sampled $3,\! 000$ points in the star's interior, uniformly distributed in ${\bm x}$ with $|{\bm x}| < R_\WD$.  The effects of this averaging are to remove the oscillatory behavior and enhance the probability at $\theta = 0, \pi$.  However, the rough approximation is reliable up to an order unity correction factor.  

Let us finally address the specific magnetic field structure in our MWD candidate, \mbox{RE J0317-853}.  Using phase-resolved far-ultraviolet spectroscopy data,~\cite{Burleigh:1998pqa} constructed several models for the field configuration of \mbox{RE J0317-853}.  It was found that the observations are well described by a magnetic dipole, with a polar field strength of $B_0 = 363 \, \mathrm{MG}$, which is offset from the star's center along the magnetic axis by $19\%$ of $R_\WD$, which is misaligned with the star's rotation axis by an angle of $20^o$, and which is viewed from an angle of $51^o$ to the rotation axis.  We have calculated the axion-photon conversion probability for this field configuration by solving \eqref{eq:matrix_eqn} and averaging over $5000$ trajectories.  Our results appear in the right panel of Fig.~\ref{fig:prob_theta}.  Since the polar viewing angle is known, we vary the azimuthal angle, $\phi$, instead.  Here we see that the axion-to-photon conversion probability experiences an $O(50\%)$ variation as the star revolves, which corresponds to a period of $\sim$$725$ s for \mbox{RE J0317-853}, and this translates into a corresponding modulation of the $X$-ray flux.  Note that across all masses we find that when integrating out in radius $r$ to compute the conversion probability the probability has researched its asymptotic values by $r \sim 10 \times R_{\rm WD}$.  

\begin{figure}[h!]
\hspace{0pt}
\vspace{-0in}
\begin{center}
\includegraphics[width=0.50\textwidth]{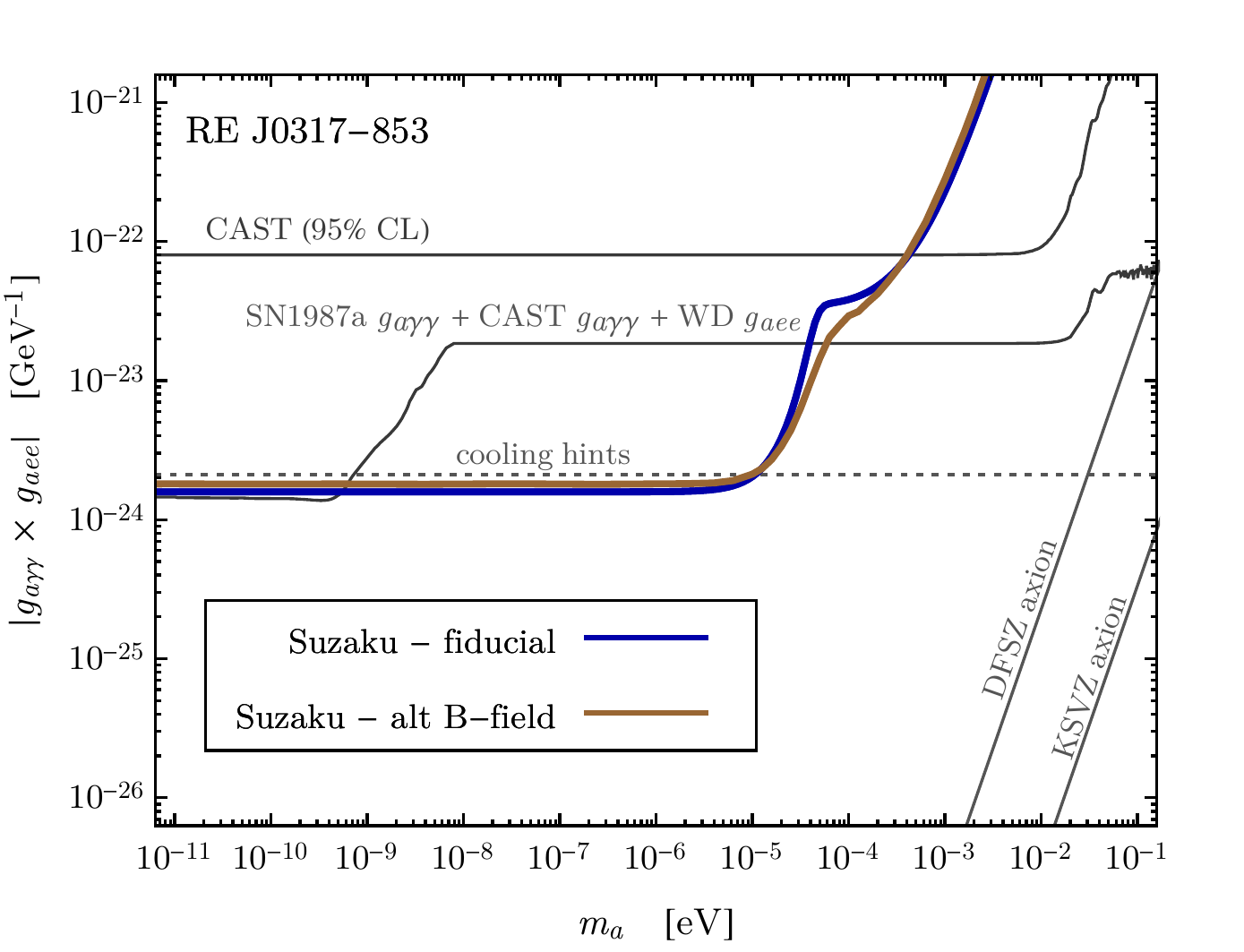}
\caption{
\label{fig:limit_w_averaging}
{
The limit on $|g_{a\gamma\gamma} \, g_{aee}|$, calculated using two different models for the background magnetic field.  The alternate $B$-field model, with the displaced dipole, gives a comparable result at low axion masses and slightly improved sensitivity at high axion masses, due to the increased magnetic field strength.  
}
}
\end{center}
\end{figure}

Finally let us compare our limit on $|g_{a\gamma\gamma} \, g_{aee}|$, calculated using two different models for the background magnetic field.  In the main text we used a configuration with $B(r) = (r/R_\WD)^{-3} \, B_0$ for $B_0 = 200 \, \mathrm{MG}$, and we obtained the limit appearing in Fig.~\ref{fig:main-result}.  If we perform the same calculation using the offset dipole configuration described above, then we obtain instead the limit appearing in Fig.~\ref{fig:limit_w_averaging} as the brown curve.  Note that the limit derived in the main text is reliable up to an $O(1)$ factor, but it is also a conservative estimate.  At small values of the axion mass, we find that the two limits are comparable.  This result is consistent with Fig.~\ref{fig:prob_theta} where we see $p_{a\to\gamma} \simeq 3.0 \times 10^{-4}$ at $\theta = \pi/2$ in the left panel while $\langle p_{a\to\gamma} \rangle \approx 2.9 \times 10^{-4}$ in the right panel after averaging over $\phi$.  We note, though, that improved sensitivity may be obtained by using an analysis procedure that incorporates the expected non-trivial light curve, in the case of the displaced dipole model.

\subsection{Effective temperature uncertainties}  
\begin{figure}[h!]
\hspace{0pt}
\vspace{-0in}
\begin{center}
\includegraphics[width=0.50\textwidth]{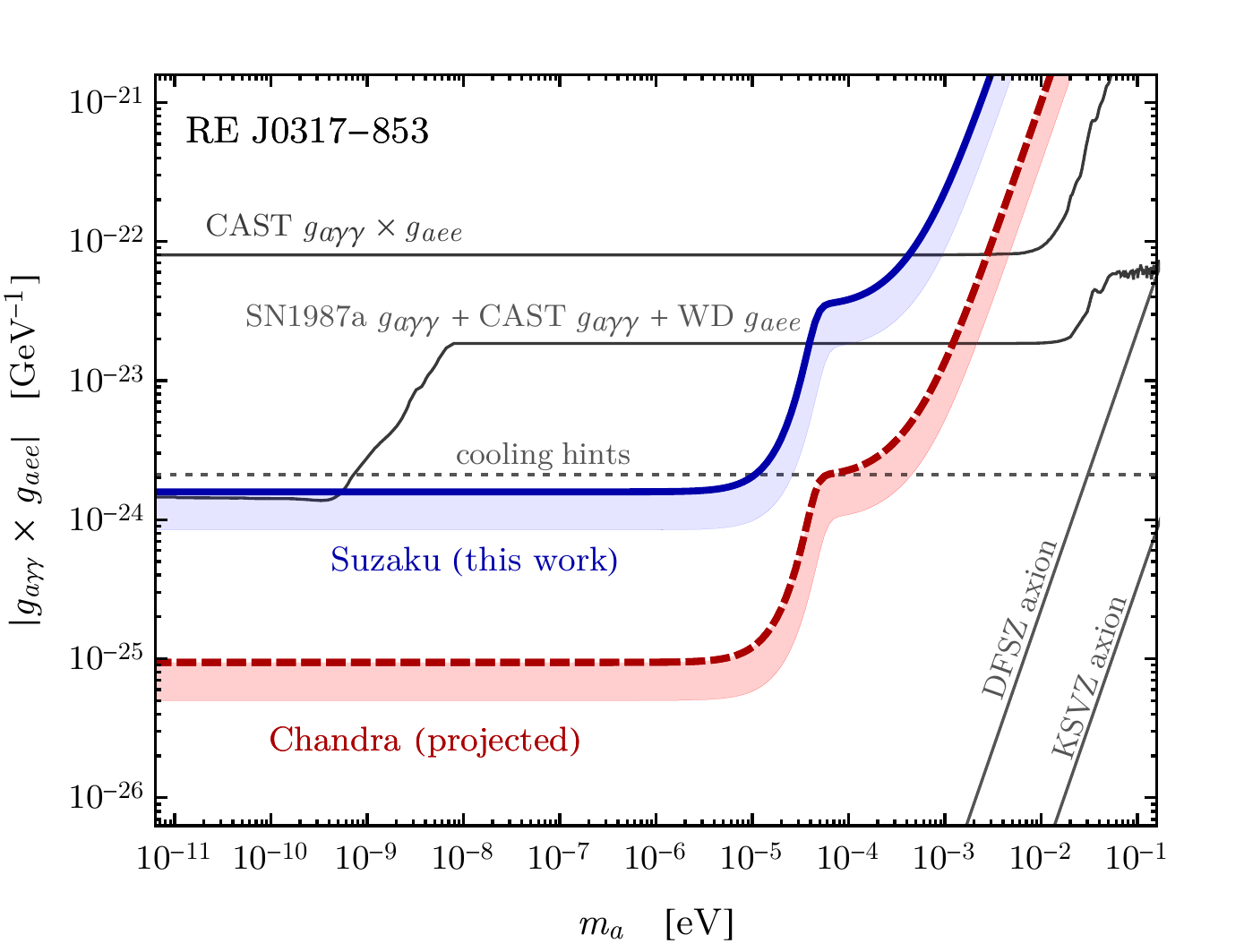}
\caption{
\label{fig: systematic}
As in Fig.~\ref{fig:main-result}, except we have broadened the {\it Suzaku} and (projected) {\it Chandra} limits to encompass the systematic uncertainty that follows from the uncertainty in the WD parameters, such as temperature, and as described in Tab.~\ref{tab:vary_Teff}.  
}
\end{center}
\end{figure}

The MWD \mbox{RE J0317-853} has garnered significant interest, due in part to its high effective temperature, which takes a value in the range from $30,\! 000$ to $55,\! 000$ K~\cite{1995MNRAS.277..971B}.  This uncertainty in $T_\mathrm{eff}$ translates into an uncertainty in our predicted $X$-ray flux, $F_{2-10}$, and therefore an uncertainty in our constraint on the axion's parameter space.  In order to quantify this uncertainty we compare our fiducial model from the main text with four benchmark models, given by Table~6 of~\cite{2010A&A...524A..36K}.  The results are summarized in Tab.~\ref{tab:vary_Teff}, which shows that the predicted $X$-ray flux can vary by a factor of $\sim 4$ depending on the specific stellar model parameters used to describe \mbox{RE J0317-853}.  Since the flux depends on the axion's couplings through $F_{2-10} \propto (g_{a\gamma\gamma} g_{aee})^2$, the effect on our limit is a factor of $\sim 2$.  The flux uncertainty follows primarily from the uncertainty in $T_\mathrm{eff}$.  In Fig.~\ref{fig: systematic} we show how these uncertainties translate into the uncertainty in the 95\% limit from {\it Suzaku} observations and (projected) {\it Chandra} observations.  The bands encompass the range of limits obtained by cycling through the parameters presented in Tab.~\ref{tab:vary_Teff}.

\begin{table}[t]
\caption{\label{tab:vary_Teff}The $X$-ray flux of \mbox{RE J0317-853} is calculated for different values of $M_\WD$, $R_\WD$, and $T_\mathrm{eff}$, which were determined by~\cite{2010A&A...524A..36K}.  The luminosities, $L_\gamma$, are inferred from the Stefan-Boltzmann law.  We calculate $F_{2-10}$ for $m_a = 10^{-9} \, \mathrm{eV}$ and $g_{a\gamma\gamma} g_{aee} = 10^{-24} \ \mathrm{GeV}^{-1}$. 
}
\begin{ruledtabular}
\begin{tabular}{c|cllcccc}
& $M_\WD \ [M_\odot]$ & $R_\WD \ [0.01 \, R_\odot]$ & $L_\gamma \ [L_\odot]$ & $T_\mathrm{eff} \ [\mathrm{K}]$ & $B \ [\mathrm{MG}]$ & $d_\WD \ [\mathrm{pc}]$ & $F_{2-10} \ [\mathrm{erg}/\mathrm{cm}^2/\mathrm{sec}]$ \\ \hline
\mbox{CO-low-T} & $1.32 \pm 0.020$ & $0.405 \pm 0.011$ & $0.0120$ & $30000$ & $200$ & $29.54$ & $6.8 \times 10^{-14}$ \\ 
\mbox{CO-high-T} & $> 1.46$ & $0.299 \pm 0.008$ & $0.0503$ & $50000$ & $200$ & $29.54$ & $2.4 \times 10^{-13}$ \\ 
\mbox{ONe-low-T} & $1.28 \pm 0.015$ & $0.416 \pm 0.011$ & $0.0126$ & $30000$ & $200$ & $29.54$ & $7.2 \times 10^{-14}$ \\  
\mbox{ONe-high-T} & $1.38 \pm 0.020$ & $0.293 \pm 0.008$ & $0.0483$ & $50000$ & $200$ & $29.54$ & $2.2 \times 10^{-13}$ \\  
\end{tabular}
\end{ruledtabular}
\end{table}

\subsection{Core temperature model uncertainties}

Since the WD's core temperature is not directly observable, it is necessary to have a model for the WD interior in order to relate $T_c$ to an observable quantity like the luminosity, $L_\gamma$.  Moreover, since the axion emission depends sensitively on the WD core temperature, with the axion luminosity going as $L_a \sim T_c^4$ in \eqref{eq:L_a}, it is therefore important to quantitatively assess how our prediction for the axion-induced X-ray signal depends on this modeling.  In this section we specifically investigate how the modeling uncertainties affect our limits on the axion parameter space for \mbox{RE J0317-853}.  

\begin{figure}[h!]
\hspace{0pt}
\vspace{-0in}
\begin{center}
\includegraphics[width=0.45\textwidth]{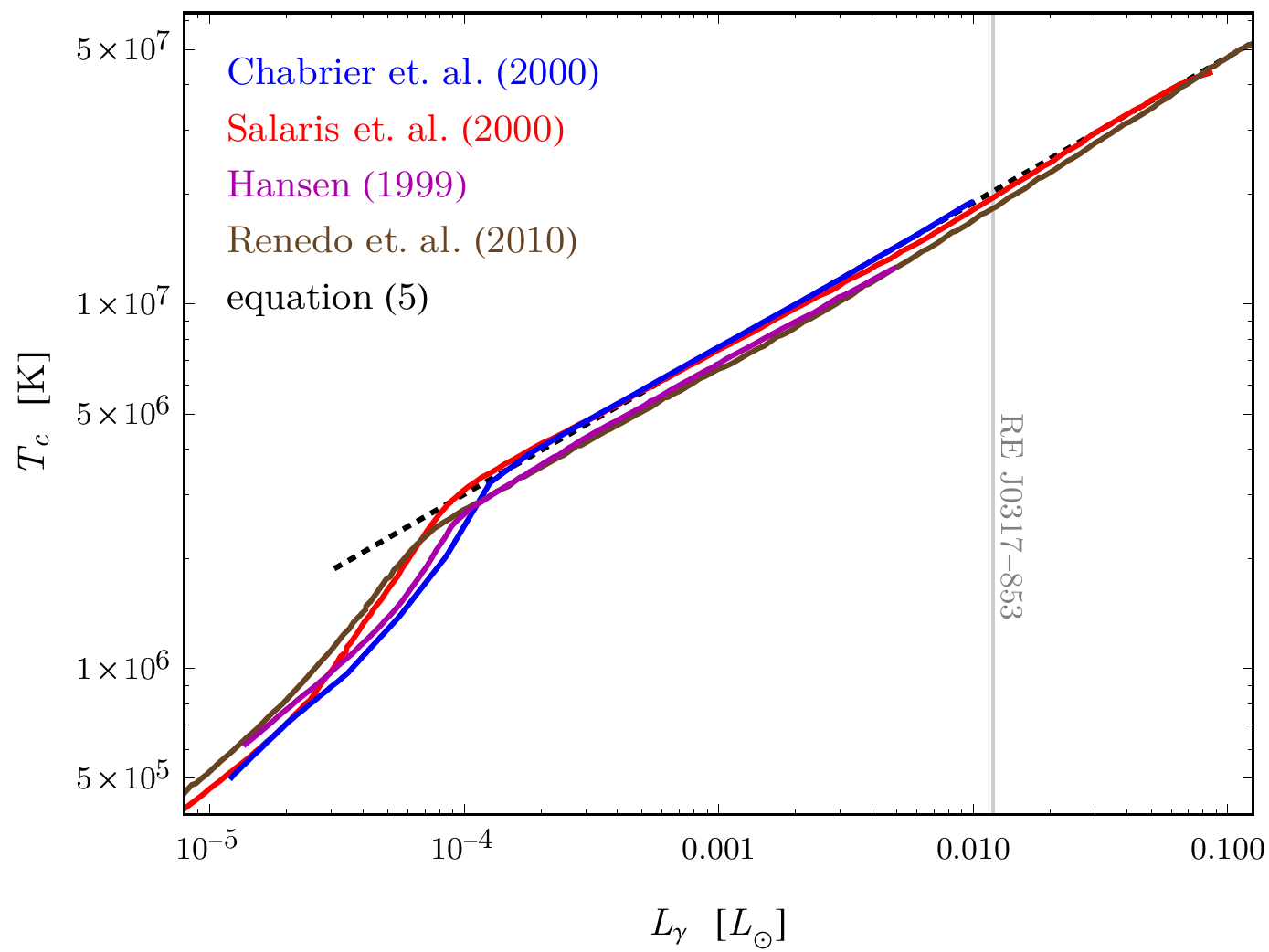} \hfill
\includegraphics[width=0.47\textwidth]{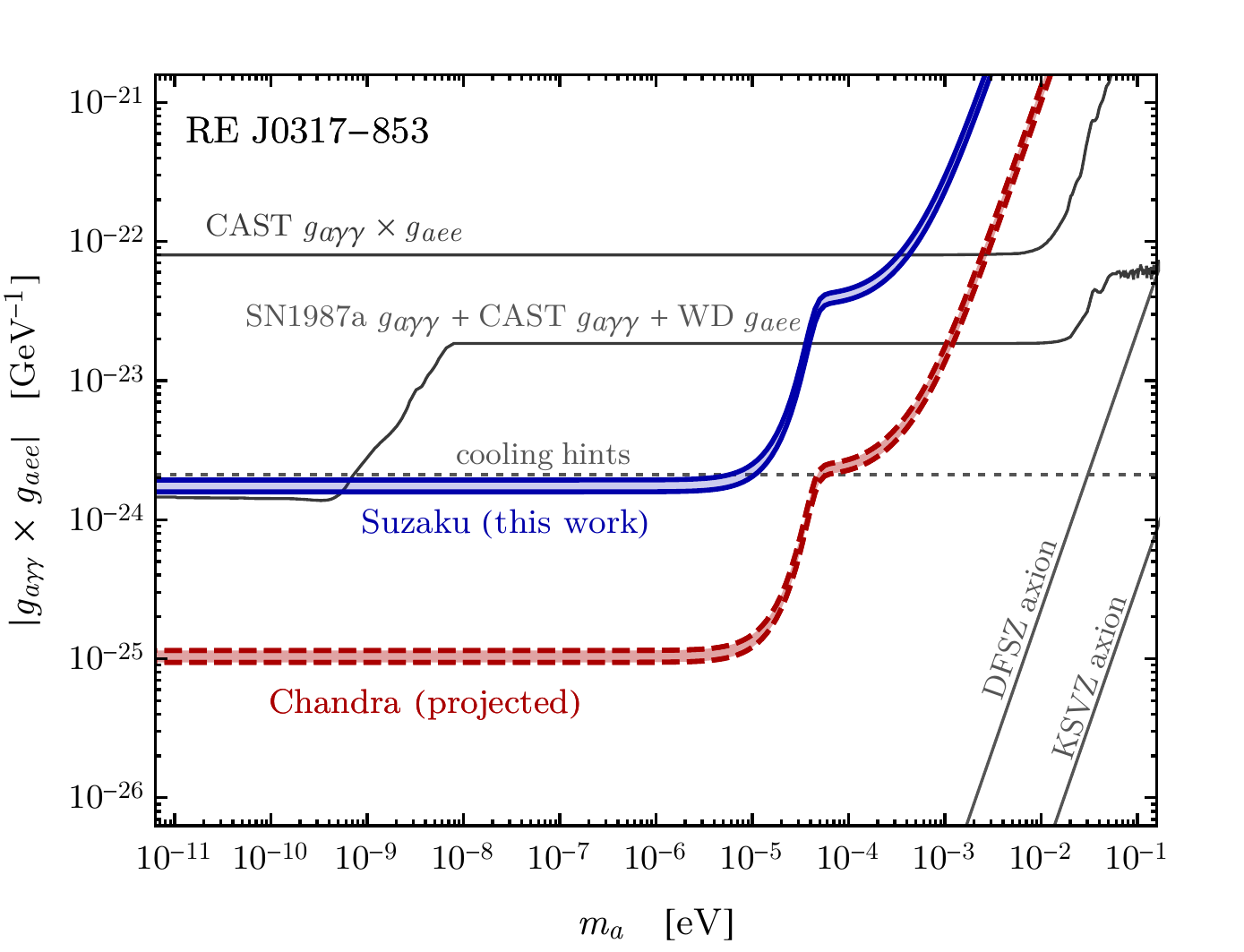}
\caption{
\label{fig:model_Tc}
A quantitative assessment of the robustness of our results under various WD cooling models.  {\it Left:}  The models in \cite{Hansen:1999fa,Salaris:2000nm,Chabrier:2000ib,Renedo:2010vb} predict the WD core temperature, $T_c$, in terms of its photon luminosity, $L_\gamma$.  Additionally the black-dashed line shows \eqref{eq:Tc_relation} and the vertical gray line indicates the fiducial luminosity for \mbox{RE J0317-853}.  {\it Right:}  As in Fig.~\ref{fig:main-result}, except that we have broadened the limit curves to reflect the uncertainty in the WD model that we use to infer $T_c$ from the measured $L_\gamma$ for \mbox{RE J0317-853}.  
}
\end{center}
\end{figure}

The physics of WD cooling is a balance between the energy stored in the hot core and the transparency of the envelope.  The transport of thermal energy also depends on various physical inputs including the thermal conductivity of the degenerate matter, neutrino emission rates, and chemical diffusion.  The cooling of old hydrogen-rich DA WDs has been studied extensively, and we summarize the relevant results of several prominent studies \cite{Hansen:1999fa,Salaris:2000nm,Chabrier:2000ib,Renedo:2010vb} in Fig.~\ref{fig:model_Tc}.  In particular we are interested in the predicted relation between the core temperature and surface luminosity.  The left panel of Fig.~\ref{fig:model_Tc} shows that all four models agree very well with our fiducial formula \eqref{eq:Tc_relation} for $10^{-4} \lesssim L_\gamma / L_\odot \lesssim 10^{-1}$.  

At the fiducial luminosity of \mbox{RE J0317-853}, $L_\gamma = 0.012 \ L_\odot$, the predictions for the core temperature vary from $T_c \simeq 1.8$ to $2.0 \times 10^7 \ \mathrm{K}$, and thus we can associate  the WD cooling model with an $O(10\%)$ uncertainty on $T_c$.  Since the axion luminosity goes as $L_a \sim T_c^4$, we infer that the WD cooling model leads to an $O(40\%)$ uncertainty on our axion-induced X-ray flux signal and an $O(20\%)$ uncertainty on our axion parameter space limits for \mbox{RE J0317-853} since $|g_{a\gamma\gamma} \times g_{aee}| \sim T_c^2$.  This is illustrated in the right panel of Fig.~\ref{fig:model_Tc}.

\subsection{Spectrum}  
\begin{figure}[h!]
\hspace{0pt}
\vspace{-0in}
\begin{center}
\includegraphics[width=0.50\textwidth]{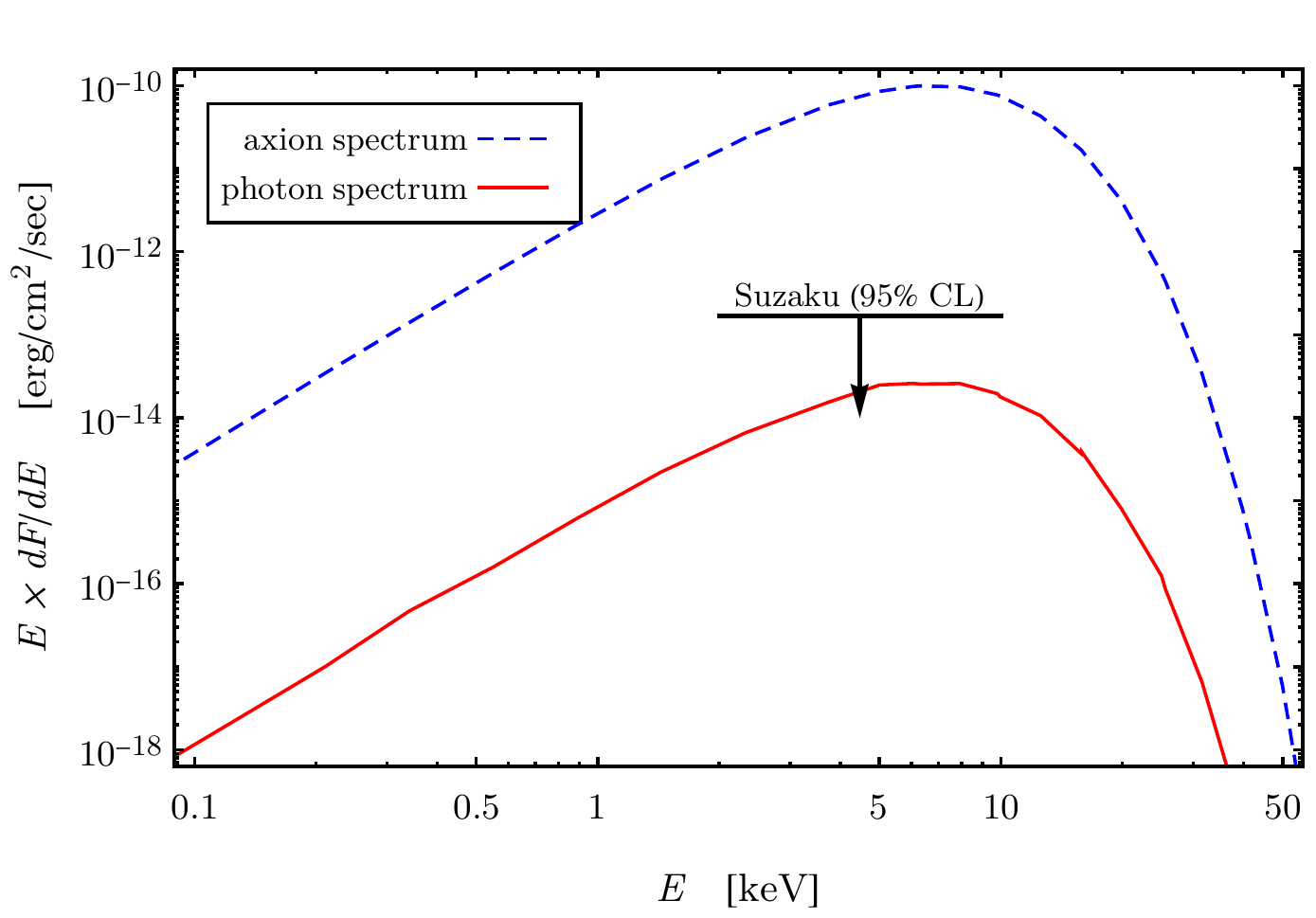}
\caption{
\label{fig:spectrum}
The predicted emission spectra of primary axions (dashed blue) and secondary $X$-rays (red) from \mbox{RE J0317-853} with $g_{aee} = 10^{-13}$, $m_a = 10^{-9} \, \mathrm{eV}$ and $g_{a\gamma\gamma} = 10^{-11} \, \mathrm{GeV}^{-1}$.  
}
\end{center}
\end{figure}

Fig.~\ref{fig:spectrum} shows the predicted spectra of axion and photon emission from our candidate MWD star, \mbox{RE J0317-853}.  The shape of the axion spectrum (blue curve) is very well approximated by a blackbody at temperature $T_c \simeq 2 \times 10^7 \, \mathrm{K} \simeq 1.7 \, \mathrm{keV}$ whereas the amplitude of the spectrum is set by the magnitude of the axion-electron coupling according to \eqref{eq:L_a}.  To draw the blue curve we take $g_{aee} = 10^{-13}$.  The spectrum of secondary, axion-induced photons (red curve) tracks the thermal spectral shape up to an additional energy dependence coming from the axion-photon conversion probability \eqref{eq:pert-prob}.  To draw the red curve we take $m_a = 10^{-9} \, \mathrm{eV}$ and $g_{a\gamma\gamma} = 10^{-11} \, \mathrm{GeV}^{-1}$, such that the product $|g_{a\gamma\gamma} \, g_{aee}| = 10^{-24}$ is marginally consistent with the {\it Suzaku} limit.  We assume that all axions are emitted isotropically and homogeneously from throughout the interior of the WD core, and the axion-photon conversion probability is calculated using \eqref{eq:matrix_eqn}.  

\section{The running of the axion-electron coupling and sensitivity to $g_{a\gamma\gamma}$}\label{SM:gaee}

In the main text we have shown the limit on $|g_{a\gamma\gamma} \, g_{aee}|$ that follows from $X$-ray observations of MWD stars.  Here we would like to translate this into a limit on simply $g_{a\gamma\gamma}$.  For any value of $g_{a\gamma\gamma}$ there is a ``reasonable range'' of values for $g_{aee}$:  it is bounded from above by direct observation, and it is bounded from below because an axion-electron coupling can be induced radiatively from an axion-photon coupling.  In this section we estimate the lower bound.  

Consider an axion-like particle, which does not couple to QCD.  The effective theory describing the interactions of this axion with the Standard Model electroweak gauge fields is generated at a high scale, which we denote by $\Lambda \sim f_a \sim 10^{10}$ GeV, and the corresponding Lagrangian can be written as 
\es{lag-high-scale}{
{\mathcal L_{a, \Lambda}} \supset -{1 \over 4} g_{aWW}^\Lambda a W \tilde W - {1 \over 4} g_{aBB}^\Lambda a B \tilde B  + {C_{e}^{\Lambda} \over 2} {\partial_\mu a \over f_a} \bar e \gamma^\mu \gamma_5 e \,,
}
where $W \equiv W^a_{\mu \nu}$ is the $SU(2)_{L}$ field strength tensor, $B \equiv B_{\mu \nu}$ is the $U(1)_Y$ field strength tensor, $g_{aWW}^\Lambda  = C_W^\Lambda \alpha_W / (2 \pi f_a)$, and $g_{aBB}^\Lambda = C_B^\Lambda \alpha_B / (2 \pi f_a)$.  The fine structure constants can be written as $\alpha_W = \alpha_\EM / s_w^2$ and $\alpha_B = \alpha_\EM / c_w^2$ where the electromagnetic fine structure constant is $\alpha_\EM \simeq 1/137$ and the weak mixing angle's sine and cosine are $s_w = \sin \theta_w \simeq 0.48$ and $c_w = \cos \theta_w \simeq 0.88$.  The parameters $C_W^\Lambda$, $C_B^\Lambda$, and $C_e^\Lambda$ are dimensionless constants that parameterize the UV theory.  We are interested in how these parameters evolve, under the renormalization group, to scales $\mu \ll \Lambda$, since the physical processes we consider are at much lower energies.  We assume that the scale $\mu$ is larger than the electroweak scale $\Lambda_{\rm EW} \sim 100$ GeV so that the Lagrangian~\eqref{lag-high-scale} is the correct description of the axion-gauge-boson interactions.  Below this scale we should instead map to the axion-photon coupling, and we will return to this point shortly.  

The one-loop diagrams that contribute to the running of $C_e$ were computed in~\cite{Srednicki:1985xd,Bauer:2017ris} and consist of triangle diagrams with electron and axion final states connected by a loop of $SU(2)_L$ or $U(1)_Y$ gauge bosons.  Evaluating the divergent part of these diagrams and applying the renormalization procedure gives the beta function for the axion-electron coupling.  The running coupling at the scale $\mu$, denoted by $C_e^\mu$, is given by~\cite{Bauer:2017ris}
\es{ce-mu}{
C_e^\mu = C_e^\Lambda + {3 \over 8 \pi^2} \alpha_\EM^2 \left( {3 \over 8} {C_W^\Lambda \over s_w^4}  + {5 \over 8} {C_B^\Lambda \over c_w^4} \right) \log {\Lambda^2 \over \mu^2} \,,
}
which illustrates how the axion-electron coupling is induced radiatively from the axion-gauge couplings in \eqref{lag-high-scale}.  Evaluating \eqref{ce-mu} at the electroweak scale, $\mu = \Lambda_{\rm EW} \simeq 100 \ \mathrm{GeV}$ gives 
\es{ce-final-old}{
C_e^\mu \approx C_e^\Lambda + (5.2 \times 10^{-4}) \, C_W^\Lambda + (7.9 \times 10^{-5}) \, C_B^\Lambda \,,
}
for typical values of $\Lambda \sim 10^{10} \ \mathrm{GeV}$.  

Below the scale of electroweak symmetry breaking we are interested in the axion-photon coupling, $\mathcal{L} \supset - g_{a\gamma\gamma} a F \tilde{F} / 4$ with $g_{a\gamma\gamma} = C_\gamma \alpha_\EM / (2\pi f_a)$.  To leading order we have $C_\gamma = C_W^\Lambda + C_B^\Lambda$, which illustrates how the axion-photon interaction arises at low energies from the axion-$W_\mu^a$ and/or axion-$B_\mu$ interactions in the UV.  Thus, barring any accidental cancellations between the $C_W$ and $C_B$ terms, we anticipate a relation between the axion-electron and axion-photon couplings, which is 
\es{ce-final}{
	|C_e^\mu| \sim (5 \times 10^{-4}) \, |C_\gamma|
}
in the IR for theories where the axion-electron coupling is not present in the UV.  Running the axion-electron coupling from the weak scale down to the electron mass scale, further enhances $C_e$ by a factor of $\sim$1-2.  Additionally, the case of the QCD axion is slightly more complicated since $C_\gamma$ receives an IR contribution from pion mixing in that case.  Therefore we simply use \eqref{ce-final} for the following estimates.  

\begin{figure}[h!]
\hspace{0pt}
\vspace{-0in}
\begin{center}
\includegraphics[width=0.50\textwidth]{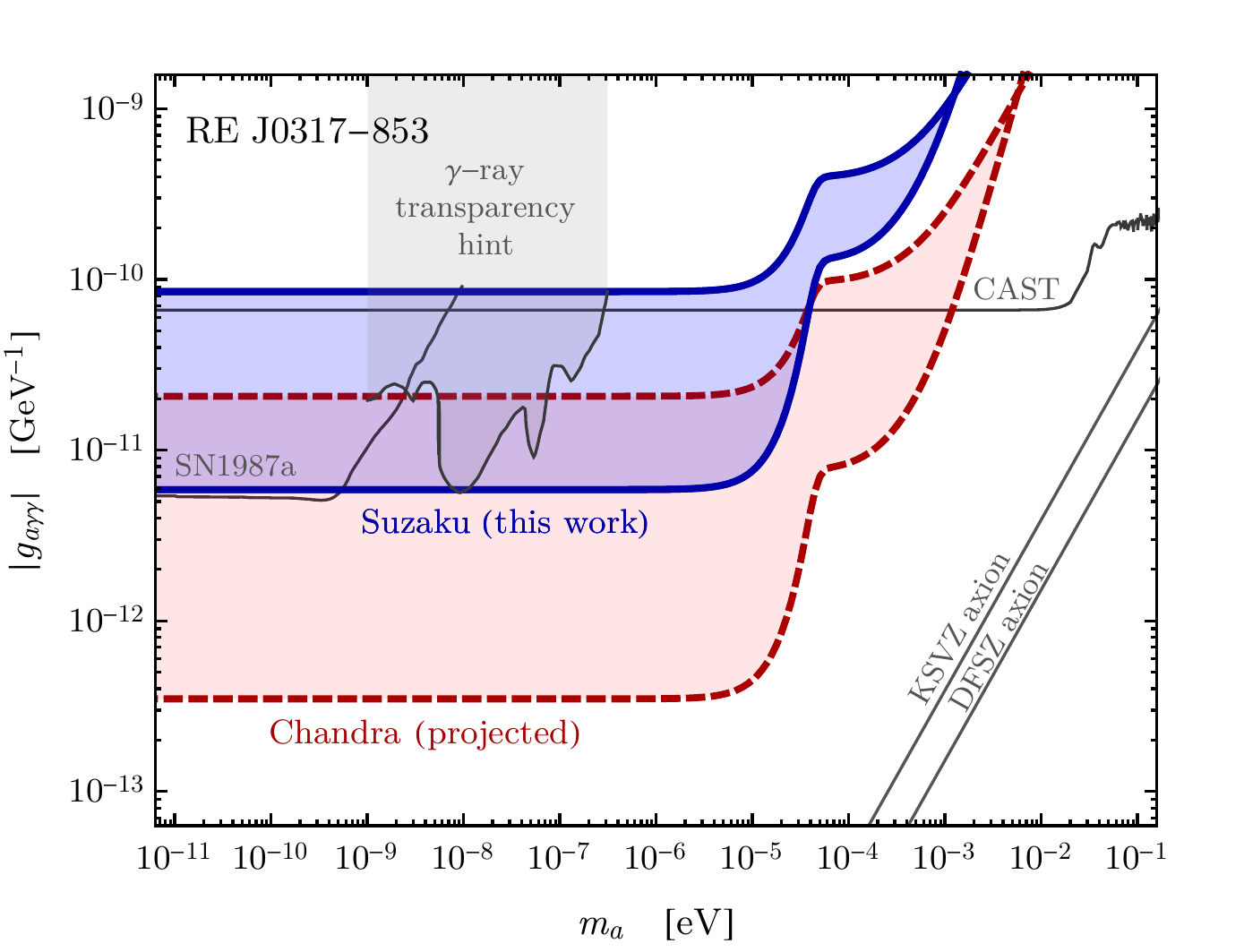}
\caption{
\label{fig:gagg_plot}
Our upper limits (from {\it Suzaku} data and projected) on $|g_{a\gamma\gamma} \, g_{aee}|$ from the main text are expressed here as upper limits on $|g_{a\gamma\gamma}|$ alone by assuming a reasonable range of values for $g_{aee}$.  For instance, the blue band is our upper limit on $|g_{a\gamma\gamma}|$ derived from {\it Suzaku} observations of \mbox{RE J0317-853}, which did not observe any $X$-ray flux.  The upper (lower) edge of the band corresponds to the smaller (larger) value of $g_{aee}$ in \eqref{gaee-range}.  
}
\end{center}
\end{figure}

In Fig.~\ref{fig:main-result} we expressed our sensitivity in terms of $|g_{a\gamma\gamma} \, g_{aee}|$, and now by using the expression for $C_e^\mu$ from \eqref{ce-final}, we can map the sensitivity onto $g_{a\gamma\gamma}$ directly.  A reasonable range of values for the axion-electron coupling is given by 
\es{gaee-range}{
	(2 \times 10^{-4}) \ |g_{a\gamma\gamma}| \ \mathrm{GeV} \ < \ |g_{aee}| \ < \ 2.8 \times 10^{-13} \,.
}
The lower limit follows from \eqref{ce-final} since $g_{aee} = C_e m_e / f_a$ and $g_{a\gamma\gamma} = \alpha_\EM C_\gamma / 2 \pi f_a$.  The upper limit is an empirical $95\%$ confidence constraint from modeling the WD luminosity function~\cite{Bertolami:2014wua}.  Using this range of values for $g_{aee}$, our sensitivity curves from Fig.~\ref{fig:main-result} translate into bands, which are shown in Fig.~\ref{fig:gagg_plot}.  We also show the CAST experiment's limit on the axion-photon coupling~\cite{Anastassopoulos:2017ftl}, and the predictions of the KSVZ and DFSZ models of the QCD axion.  For $m_a \lesssim 10^{-5} \ \mathrm{eV}$ our limit is comparable to the CAST 95\% CL exclusion, even under the ``pessimistic'' assumption that $g_{aee}$ is ``minimal'' as in \eqref{ce-final}, which leads to the weakest limit on $g_{a\gamma\gamma}$.  On the other hand, dedicated {\it Chandra} observations would lead to a conservative upper limit on $g_{a\gamma\gamma}$ that is significantly stronger than the CAST bound at low masses.  We also show an upper bound on $|g_{a\gamma\gamma}|$ from gamma-ray flux associated with SN1987a~\cite{Payez:2014xsa}.  

Our results also constrain the axion explanations of the very-high-energy gamma-ray transparency anomalies previously observed with Cherenkov telescope data~\cite{Horns:2012fx,Meyer:2013pny}.  The gamma-ray observations indicate that the Universe is more transparent than previously thought to high-energy gamma-rays, and one explanation is that gamma-rays could oscillate into axions in astrophysical magnetic fields.  The allowed parameter space from~\cite{Meyer:2013pny} to fit these anomalies is indicated in shaded gray in Fig.~\ref{fig:gagg_plot}, though we note that stringent constraints from {\it e.g.} the {\it Fermi} Large Area Telescope~\cite{TheFermi-LAT:2016zue} and H.E.S.S.~\cite{Abramowski:2013oea} also exist below $\sim$${\rm few} \times 10^{-8}$ eV.  Furthermore, the transparency anomalies are subject to uncertainties from assumptions about galactic magnetic field models.  Future dedicated observations of MWDs by {\it e.g.} {\it XMM-Newton} or {\it Chandra} would be able to probe much of the motivated parameter space to explain these anomalies.

\begin{figure}[h!]
\hspace{0pt}
\vspace{-0in}
\begin{center}
\includegraphics[width=0.50\textwidth]{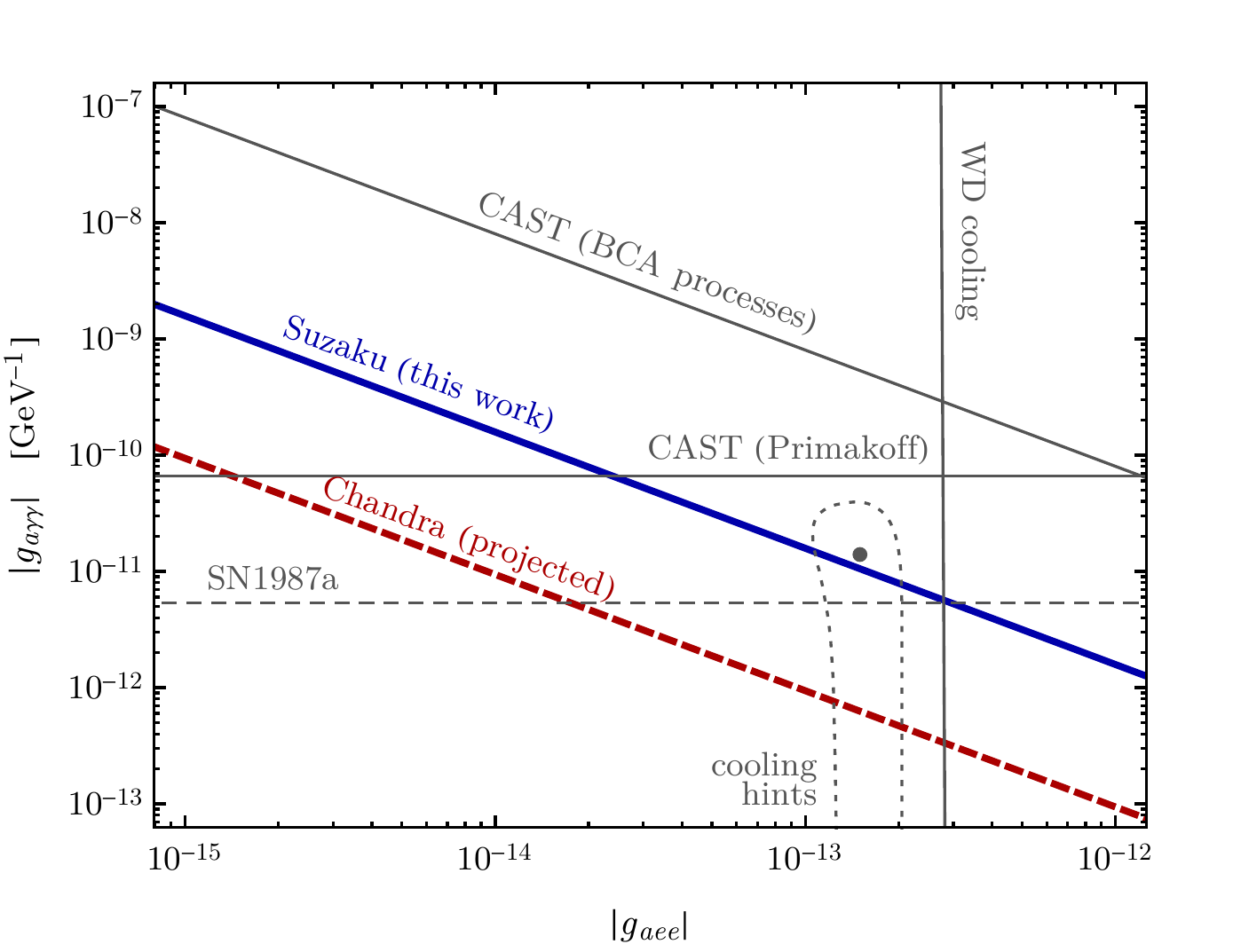}
\caption{
\label{fig:gagg_gee}
This figures summarizes upper limits on the couplings of axions with photons and electrons for low mass axions where these limits become insensitive to $m_a$.  The CAST helioscope provides both an upper limit on $|g_{a\gamma\gamma}|$ from axions produced in the Sun through the Primakoff process, as well as an upper limit on $|g_{a\gamma\gamma} g_{aee}|$ from axions produced through the BCA processes:  bremstrahlung, Compton, and axio-recombination.  We also highlight the region of parameter space that is favored by the various stellar cooling hints~~\cite{Giannotti:2017hny} with the best-fit point indicated with a gray dot, and the $1\sigma$ confidence region indicated by a gray-dotted curve.  Observations of SN1987a~\cite{Payez:2014xsa} imply an upper limit on $|g_{a\gamma\gamma}|$ for $m_a < 10^{-9} \ \mathrm{eV}$ at the level shown by the gray-dashed line, but this limit becomes weaker than the CAST limit above $m_a \approx 10^{-8} \ \mathrm{eV}$.  
}
\end{center}
\end{figure}

For low-mass axions with $m_a \lesssim 10^{-6} \ \mathrm{eV}$, the upper limits on axion-matter couplings become insensitive to the axion's mass, $m_a$.  We summarize these limits in Fig.~\ref{fig:gagg_gee}, where we also compare our limit with previous limits on the axion's coupling to photons, $g_{a\gamma\gamma}$ and the axion's coupling to electrons, $g_{aee}$.  

\end{document}